\documentclass[pdflatex,sn-mathphys-num]{sn-jnl}

\usepackage{graphicx}%
\usepackage{multirow}%
\usepackage{amsmath,amssymb,amsfonts}%
\usepackage{amsthm}%
\usepackage{mathrsfs}%
\usepackage[title]{appendix}%
\usepackage{xcolor}%
\usepackage{textcomp}%
\usepackage{manyfoot}%
\usepackage{booktabs}%
\usepackage{algorithm}%
\usepackage{algorithmicx}%
\usepackage{algpseudocode}%
\usepackage{listings}%
\usepackage{lineno}
\usepackage{hyperref}
\usepackage[table]{xcolor}   
\usepackage[normalem]{ulem}

\definecolor{eocol}{HTML}{EEF2FB}       
\definecolor{piezocol}{HTML}{E8F5E9}    
\definecolor{thiswork}{HTML}{FCE4D6}    
\definecolor{memscol}{HTML}{FFF8E7}     
\definecolor{thermocol}{HTML}{FDE8E8}   
\definecolor{aocol}{HTML}{E3F4FB}       
\definecolor{lccol}{HTML}{F3F0FB}       
\definecolor{sechead}{HTML}{D9D9D9}     

\theoremstyle{thmstyleone}%
\theoremstyle{thmstyletwo}%

\theoremstyle{thmstylethree}%

\begin{document}

\title[Article Title]{Ultra-low loss piezo-optomechanical low-confinement silicon nitride platform for visible wavelength quantum photonic circuits}

\author[1]{\fnm{Mayank} \sur{Mishra}}

\author[1]{\fnm{Gwangho} \sur{Choi}}

\author[1]{\fnm{Wenhua} \sur{He}}

\author[2]{\fnm{Gina M.} \sur{Talcott}}

\author[2]{\fnm{Katherine} \sur{Kearney}}

\author[2]{\fnm{Michael} \sur{Gehl}}

\author[2]{\fnm{Andrew} \sur{Leenheer}}

\author[2]{\fnm{Daniel} \sur{Dominguez}}

\author*[2]{\fnm{Nils T.} \sur{Otterstrom}}\email{ntotter@sandia.gov}

\author*[1,2,3]{\fnm{Matt} \sur{Eichenfield}}\email{Matt.Eichenfield@colorado.edu}

\affil*[1]{\orgdiv{James C. Wyant College of Optical Sciences}, \orgname{University of Arizona}, \orgaddress{\city{Tucson}, \state{Arizona}, \postcode{85721}, \country{USA}}}

\affil[2]{\orgdiv{Microsystems Engineering, Science, and Applications}, \orgname{Sandia National Laboratories}, \orgaddress{\city{Albuquerque}, \state{New Mexico}, \postcode{87185}, \country{USA}}}

\affil[3]{\orgdiv{Electrical, Computer and Energy Engineering}, \orgname{University of Colorado}, \orgaddress{\city{Boulder}, \state{Colorado}, \postcode{80309}, \country{USA}}}

\abstract{Realizing photonic quantum computing at scale requires integrated circuits that combine ultra-low loss with fast, low-power, low-hysteresis, and low-crosstalk reconfiguration. These requirements are particularly challenging at visible wavelengths, where many quantum resource-state generators, including single-photon sources and quantum memories, naturally operate. Low-confinement silicon nitride waveguides offer the requisite loss performance, but conventional thermo-optic modulators dissipate significant static power, driving thermal crosstalk and precluding cryogenic operation. Visible-wavelength piezo-optomechanical circuits avoid these drawbacks, but existing demonstrations rely either on high-confinement waveguides with propagation losses of 35–100 dB/m, or on low-confinement platforms using foundry-incompatible PZT. Here we combine piezo-optomechanical actuation with a CMOS-foundry-fabricated, low-confinement silicon nitride platform, achieving 2.6 dB/m propagation loss at 780 nm, megahertz-scale modulation bandwidth, a half-wave voltage-length product of 2.8 V·m, and negligible hysteresis. We demonstrate reconfigurable Mach–Zehnder interferometers with 0.63 dB loss per spiral phase shifter, enabling deep, actively reconfigurable visible-wavelength quantum photonic circuits.}


\maketitle

\section*{Introduction}\label{sec1}

Recent work in high-confinement, piezo-optomechanical photonic integrated circuits (PICs) in the visible wavelength regime has demonstrated high circuit densities, fast modulation rates, and low power dissipation/consumption \cite{Dong22_01, Dong22_05, Dong23_04, Palm23_05, Stanfield19_09, clark23_08}. These capabilities make piezo-optomechanical PICs well suited for realizing linear programmable nanophotonic processors (LPNPs)  \cite{Harris18_12}, which are scalable, reconfigurable photonic circuits that implement arbitrary linear optical transformations, enabling applications such as quantum resource state generation and linear optical quantum computing. Given this set of capabilities, piezo-optomechanical PICs could be well-suited for visible wavelength scalable quantum photonic information processing applications such as integrated quantum resource state generators \cite{bhatti2023GHZwithSMS}, linear optical computing \cite{optical_comput}, hybrid spin-photon quantum computing architectures \cite{clark23_08}, quantum networking protocols \cite{Lee2022}, and quantum sensors \cite{Sensing_18}. However, the propagation losses in these high-confinement waveguides remain, in many cases, too high to achieve the circuit depths and low error rates necessary for scalable quantum computing \cite{rudolph2017optimistic}. In contrast, low-confinement silicon nitride ($\mathrm{Si}_{3}\mathrm{N}_4$) waveguides, which feature high-aspect ratio silicon nitride cores and thick silicon dioxide claddings, have the potential to address this scaling challenge by offering ultra-low propagation losses \cite{Bauters_11, Hong_22}. These waveguides have shown ultra-low losses of less than 0.1 dB/m \cite{Bauters11_09, Jin2021} at telecommunication wavelengths and below 1 dB/m \cite{Chauhan22, Chauhan02_22} at visible wavelengths. These low-confinement waveguide platforms deliver high optical power-handling capabilities and a wide transparency window \cite{Blumenthal20_02, Blumenthal18}. Furthermore, they support the realization of ultra-high-Q resonators \cite{Isichenko2024} and provide a pathway for integration with quantum systems, including rubidium atoms \cite{Isichenko2023}, strontium atoms \cite{Chauhan2021}, and InAs quantum dots \cite{NatCom:22}.

The integration of thermo-optic modulators \cite{Yong:22, Isichenko2024, Liu:24} and PZT-based piezo-optomechanical actuators \cite{Hosseini:15, Wang:221, nick26_Blumenthal} with low-confinement silicon nitride waveguides has been demonstrated, but both approaches exhibit significant limitations. Thermo-optic modulators typically exhibit slow modulation speeds, high power consumption/dissipation, and large crosstalk. On the other hand, PZT-based devices offer higher modulation speed and lower crosstalk but suffer from significant ferroelectric hysteresis and integration limitations due to CMOS-incompatibility. In contrast, piezo-optomechanical actuator platforms based on aluminum nitride (AlN) implemented using CMOS foundry processes offer a promising alternative \cite{Tian:241, Dong22_01, Dong22_05, Dong23_04, Stanfield19_09}. Although so far only demonstrated with high-confinement waveguides, these AlN actuators provide high modulation bandwidths (exceeding $100~\mathrm{MHz}$), ultralow power consumption with negligible hysteresis, and compatibility with cryogenic operation.

\begin{figure}
\centering
\includegraphics[width=\linewidth]{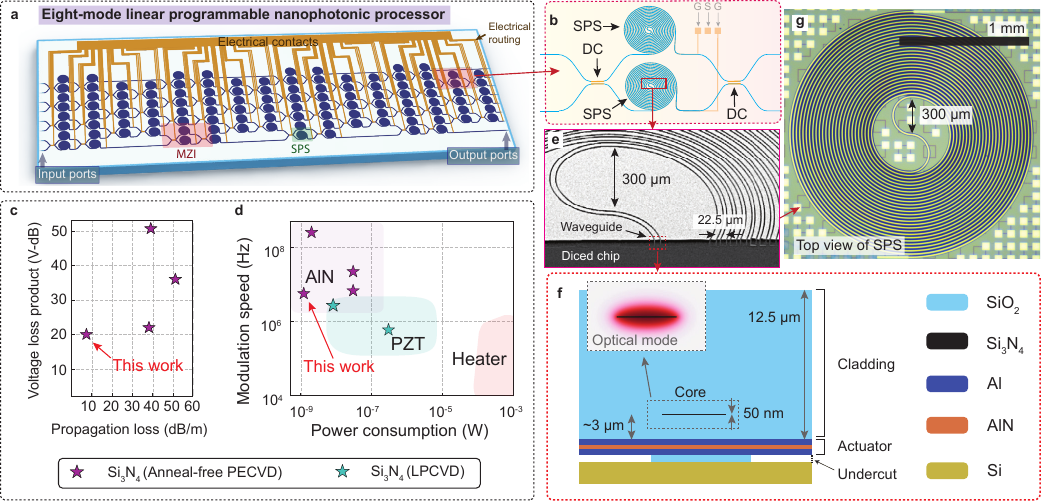}
\caption{\textbf{Low-confinement silicon nitride piezo-optomechanical platform.}\\
\textbf{(a)} Schematic representation of the eight-mode linear programmable nanophotonic processor (LPNP) circuit, based on the Clements et al. \cite{Clements16} architecture, featuring a network of Mach-Zehnder interferometers (MZIs) and phase shifters. \textbf{(b)} Schematic of an MZI incorporating two Archimedean spiral phase shifters (SPSs) and two tunable directional couplers (TDCs). \textbf{(c)} Comparison of the voltage-loss product ($\mathrm{VLP}$) and propagation loss across existing visible wavelength aluminum nitride-based piezo-optomechanical modulator platforms \cite{Dong22_01,Dong22_05}. \textbf{(d)} Comparison of modulation speed and power consumption for visible wavelength silicon nitride ($\mathrm{Si}_3\mathrm{N}_4$) based modulator platforms \cite{Dong22_01, Dong22_05, Hosseini:15, Yong:22, nick26_Blumenthal}. The $\mathrm{Si}_3\mathrm{N}_4$ platforms include devices fabricated via low-pressure chemical vapor deposition (LPCVD) and anneal-free plasma-enhanced chemical vapor deposition (PECVD) \cite{Bose2024}. Details on the loss and $\mathrm{VLP}$ values for existing visible wavelength $\mathrm{Si}_3\mathrm{N}_4$ platforms are provided in Supplementary Table 1. \textbf{(e)} Scanning electron microscope (SEM) image showing the cross-section of a wafer diced through the center of an on-chip SPS, with a minimum bending radius of $150~\mu\mathrm{m}$ at the center. \textbf{(f)} Cross-sectional schematic of a low-confinement waveguide-based piezo-optomechanical phase modulator, highlighting the different material layers. Inset: x-component of the electric field of the fundamental optical mode in a low-confinement waveguide, simulated using the finite element method. \textbf{(g)} Top-view micrograph of the $9.8~\mathrm{cm}$-long spiral phase shifter, featuring a minimum bend radius of $150~\mu\mathrm{m}$, a footprint of $3.80~\mathrm{mm}^2$ ($1.95~\mathrm{mm} \times 1.95~\mathrm{mm}$), and a $22.5~\mu\mathrm{m}$ gap between adjacent spiral waveguides.}
\label{Fig1}
\end{figure}

 The much thicker silicon dioxide claddings required by low-confinement silicon nitride waveguides present a key challenge for piezo-optomechanical actuation and strain-optic phase shifters. Without careful design, the increased oxide thickness leads to high mechanical rigidity and, consequently, reduced piezo‑optomechanical responsivity. In this work, we show that thick silicon dioxide claddings with an extended undercut beneath the actuator can enable geometries and aspect ratios that concentrate the distribution of stress and optical field within the core and cladding of the waveguide in a way that allows high piezo-optomechanical responsivity even with thick claddings. Moreover, because a large fraction of the optical field in these low‑confinement waveguides resides in the silicon dioxide cladding, the comparatively stronger photoelastic response of silicon dioxide \cite{PhysRevLett.124.013902, Tian:241}, despite its lower refractive index than silicon nitride, compensates for the reduced mechanical compliance (see Supplementary Fig.~3). Experimentally, we demonstrate a design achieving piezo-optomechanical responsivity comparable to high-confinement systems while lowering optical loss by an order of magnitude \cite{Dong22_01}. While our focus is on low-confinement visible-wavelength platforms, active modulation has been extensively demonstrated in high-confinement systems, including electro-optic modulators based on thin-film lithium niobate \cite{renaud2023sub, celik2022high, li2022high, desiatov2019ultra}, lithium tantalate \cite{powell2025sub}, and aluminum nitride \cite{xiong2012low, zhu2019aluminum}, as well as electrostatic MEMS phase shifters \cite{govdeli2025a} and thermo-optic tuners \cite{mohanty2020recon, wu2024submilliwatt, yong2022p, liang2021r}. These approaches provide a range of modulation and reconfiguration mechanisms; however, the tightly confined modes they employ generally incur higher propagation loss, limiting scalability in large photonic circuits. A quantitative comparison of these platforms with our piezo-optomechanical platform, including modulation mechanism, operating wavelength, propagation loss, $V_{\pi}L$, modulation bandwidth, insertion loss, and device footprint, is provided in Supplementary Table~2. 
 
In this work, we present a low-confinement platform that combines ultra-low propagation loss with piezo-optomechanically actuated phase modulation. We characterize the propagation loss, modulation efficiency, and voltage–loss product of our low-confinement platform, and demonstrate, through an example LPNP circuit, that cascading longer, low-loss phase shifters offers a decisive advantage over shorter, higher-loss alternatives, enabling orders-of-magnitude improvements in the success probability of heralding multi-photon quantum states. Together, these results establish low-loss phase modulators as a key enabler for large-scale programmable visible-wavelength photonic integrated circuits.

\section*{Results}

\subsection*{Low-confinement waveguide platform}
\label{sec2}

The fundamental programmable element of low-confinement waveguide linear programmable nanophotonic processor  platform (Fig.~\ref{Fig1}a) is a Mach-Zehnder interferometer (MZI; Fig.~\ref{Fig1}b) with integrated output phase shifters. These MZIs utilize Archimedean‑spiral‑shaped waveguide phase shifters to reduce both the device footprint and optical bending losses. A scanning electron micrograph of a spiral phase-shifter (SPS) section fabricated using an anneal-free plasma-enhanced chemical vapor deposition (PECVD) process~\cite{Stanfield19_09,Dong22_01,Dong22_05,Dong23_06, Bose2024} is shown in Fig.~\ref{Fig1}e, and a top-view micrograph of the SPS in Fig.~\ref{Fig1}g. Compared to conventional low-pressure chemical vapor deposition (LPCVD), the anneal-free PECVD approach enables low-temperature processing, allowing seamless monolithic integration with CMOS metal routing for the actuator beneath the waveguide, which would otherwise be damaged by high-temperature annealing. As detailed in subsequent sections and highlighted in Fig.~\ref{Fig1}c, our low-confinement waveguides establish a benchmark in voltage-loss product (VLP) and propagation loss among visible wavelength, anneal-free PECVD-fabricated silicon nitride waveguides with AlN piezo-optomechanical actuators \cite{Bose2024}. Additionally, as shown in Fig.~\ref{Fig1}d, these devices achieve high modulation speeds in the MHz range and exhibit low holding power in the nanowatt range ($P_{\mathrm{hold}} < 1~\mathrm{nW}$ for a 5 V applied voltage at room temperature), comparable to other visible wavelength modulators employing AlN-based piezo-optomechanical strain actuator platforms~\cite{Dong22_01, Dong22_05}.

\begin{figure}
\centering
\includegraphics[width=\linewidth]{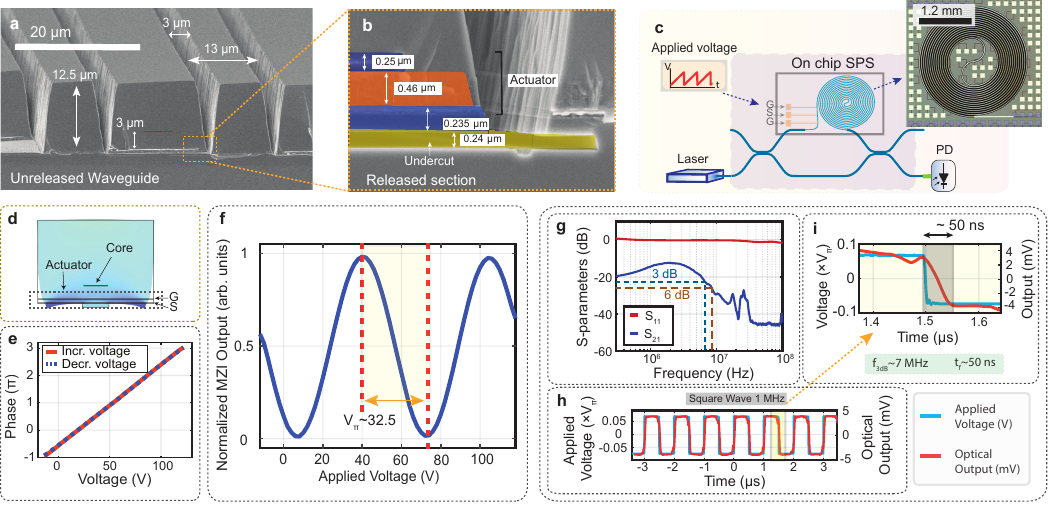}
\caption{\textbf{Piezo-optomechanical phase shifter characterization.} \\
\textbf{(a, b)} Cross-sectional SEM images of the fabricated phase modulators. In \textbf{(b)}, blue and orange false colors highlight the electrode and the AlN actuator layer, respectively, which is suspended by an undercut. \textbf{(c)} Schematic of the imbalanced MZI used for characterization, incorporating an on-chip spiral phase shifter in one arm, two off-chip 50:50 beam splitters, a 780 nm laser, and a photodetector. A ramp voltage (red) is applied to the piezo-optomechanical actuator, modulating the phase and producing a sinusoidal optical output. Inset: top-view micrograph of the $8.7~\mathrm{cm}$-long spiral phase shifter, with a minimum bend radius of $300~\mu\mathrm{m}$ and a chip footprint of 4.84 $\mathrm{mm}^2$ ($2.2~\mathrm{mm} \times 2.2~\mathrm{mm}$). \textbf{(d)} Schematic cross-section of the phase shifter with FEM-simulated $s_{11}$ strain distribution and the corresponding (exaggerated) waveguide deformation for clarity. \textbf{(e)} Measured phase response of the MZI for increasing and decreasing voltage sweeps from -10 V to 120 V and back, showing negligible hysteresis. \textbf{(f)} Measured optical response of the MZI under a $200~\mathrm{Hz}$ ramp voltage applied to the piezo-optomechanical SPS. An $8.7~\mathrm{cm}$ SPS achieves a $\pi$ phase shift at $32.5~\mathrm{V}$. \textbf{(g)} Measured S-parameter response ($S_{11}$, $S_{21}$) of the SPS. 
$S_{21}$ exhibits a 3-dB roll-off at approximately 7~MHz, while $S_{11}$ remains featureless across the 0--100~MHz measurement range. \textbf{(h)} Optical response to a 1~MHz square-wave voltage drive, showing approximately six complete periods. \textbf{(i)} Zoomed-in view of a single falling edge from \textbf{(h)}, revealing a 90--10\% fall time of approximately 50~ns.}
\label{Fig2}
\end{figure}

Fig.~\ref{Fig1}f illustrates the integration of the low-confinement waveguide with a piezo-optomechanical actuator. A scanning electron micrograph of the waveguide is shown in Fig.~\ref{Fig2}a, with the thin core highlighted in yellow (false color). The fabricated waveguide features a $50~\mathrm{nm}$-thick, $3.6~\mu\mathrm{m}$-wide core, positioned $2.96~\mu\mathrm{m}$ above the actuator, and is embedded within a $\mathrm{SiO}_2$ cladding that is $12.5~\mu\mathrm{m}$-thick and $15.8~\mu\mathrm{m}$-wide at the core height, as depicted in Fig.~\ref{Fig2}a. The actuator, shown in Fig.~\ref{Fig2}b, consists of a $0.46~\mu\mathrm{m}$-thick aluminum nitride layer sandwiched between two $0.25~\mu\mathrm{m}$-thick aluminum electrodes, connected to electrical signals through ground-signal-ground (GSG) contact pads. To characterize the SPS, we integrate it into one arm of an imbalanced MZI, as illustrated in Fig.~\ref{Fig2}c. Adiabatic tapers ($2~\mathrm{mm}$ long) at both input and output ports gradually widen the core from $0.4~\mu\mathrm{m}$ to $3.6~\mu\mathrm{m}$, ensuring fundamental mode excitation \cite{chiu2019critical} and minimizing coupling loss.

When a voltage difference $\Delta V$ is applied across the electrical contacts, it generates a vertical electric field given by $E_3 \approx \Delta V / t_{\mathrm{AlN}}$, where $t_{\mathrm{AlN}}$ is the AlN layer thickness. This electric field induces strain in the structure, described by the strain tensor $s_{kl}$, with the $s_{11}$ component illustrated in Fig.~\ref{Fig2}d. The induced strain modulates the effective refractive index ($n_{\mathrm{eff}}$) of the waveguide via two primary mechanisms: the photoelastic effect ($\Delta n_{\mathrm{eff,PE}}$) and the moving boundary effect ($\Delta n_{\mathrm{eff,MB}}$) \cite{Stanfield19_09, Tian:241}. The relative change of refractive index under strain is conventionally described by the
strain-optic coefficient as $\Delta (1/n^2)_{ij}=\sum_{k,l} p_{ijkl} s_{kl}$, where $p_{ijkl}$ is the photoelastic tensor \cite{PhysRevLett.124.013902, 10.1063/1.1709293, Tian:241}. The strain components are related to the applied electric field by
$s_{kl} = \sum_{m} d_{mkl} E_{m}$, where $d_{mkl}$ represents the piezoelectric coupling tensor.
The total change in effective refractive index per applied volt,
$\Delta n_{\mathrm{eff}} = \Delta n_{\mathrm{eff,PE}} + \Delta n_{\mathrm{eff,MB}}$,
produces a phase shift in the optical mode given by
$\Delta\phi = \frac{2\pi}{\lambda} \Delta n_{\mathrm{eff}} L$. Using two-dimensional finite element method (FEM) simulations of the waveguide cross section (as shown in Fig.~\ref{Fig2}d and the inset of Fig.~\ref{Fig1}f), we calculate the voltage-length product ($V_{\pi} L$):
\begin{equation}
V_{\pi}L
=
\frac{\lambda}{2}
\left(
\frac{\Delta n_{\mathrm{eff}}}{\Delta V}
\right)^{-1}.
\end{equation}
where $\Delta n_{\mathrm{eff}}/\Delta V$ is the change in the effective refractive index per applied volt. Supplementary Fig.~9 further analyze how $V_{\pi} L$ varies with geometry.

To characterize the modulator, we apply a $200~\mathrm{Hz}$ ramp signal to the SPS, sweeping the voltage from $-10~\mathrm{V}$ to $120~\mathrm{V}$. The resulting optical response exhibits sinusoidal fringes due to the MZI transfer function, with negligible hysteresis between the forward and reverse voltage sweeps, as shown in Fig.~\ref{Fig2}e. From these measurements, we extract an average $V_{\pi}$ of approximately $32.5~\mathrm{V}$. For the 8.7-cm-long spiral phase shifter at $\lambda = 780~\mathrm{nm}$, this corresponds to a voltage-length product of $V_{\pi}L = 2.8~\mathrm{V}\cdot\mathrm{m}$. Additional measurement details on other tested devices are provided in Supplementary Fig.~4.

These devices are tested well below any mechanical resonance frequencies. Although operating at or above resonance can enhance the optomechanical response, it also induces narrowband behavior, ringing, and phase lag, so the fundamental eigenfrequency ultimately sets the upper limit of the usable bandwidth. We characterized the modulator's frequency response using a vector network analyzer (VNA), as shown in Fig.~\ref{Fig2}g. The transmission coefficient $S_{21}$ rolls off by 3~dB at approximately $f_\mathrm{3dB} = 7~\mathrm{MHz}$, defining the modulation bandwidth of the piezo-optomechanical actuators and by 6~dB at approximately $9~\mathrm{MHz}$, while the reflection coefficient $S_{11}$ remains essentially flat, with no resonant features over the $0$--$100~\mathrm{MHz}$ range. We confirmed this bandwidth in the time domain by driving the device with a $1~\mathrm{MHz}$ square-wave voltage signal and recording the optical response (Fig.~\ref{Fig2}h), which yields a 90--10\% fall time of approximately $t_f = 50~\mathrm{ns}$ (Fig.~\ref{Fig2}i). These two independent measurements agree quantitatively: for a single-pole 
(first-order) response, they are related by $t_f \cdot f_\mathrm{3dB} \approx 0.35$, equivalent to a time constant $\tau = 1/(2\pi f_\mathrm{3dB}) \approx 23~\mathrm{ns}$. Using FEM simulations of the device, which experimentally exhibited a $V_{\pi} L$ of $2.8~\mathrm{V}\cdot\mathrm{m}$, together with reported literature values of $|p{12}|$ and $|p_{44}|$ \cite{Tian:241}, we roughly infer plausible values for the silicon nitride 
photoelastic coefficients to guide our future device designs. Because $p_{44} = \frac{p_{11} - p_{12}}{2}$, there are four candidate combinations of $p_{11}$ and $p_{12}$, and only $p_{11} = -0.125$ and $p_{12} = 0.047$ yield simulated values of $V_{\pi} L$ consistent with the experimental measurement of $2.8~\mathrm{V}\cdot\mathrm{m}$. These findings indicate that silicon dioxide dominates the strain-induced effective refractive index change, whereas silicon nitride contributes an opposing effect, as discussed in Supplementary Note 3.

\subsection*{Loss characterization of low-confinement waveguide}
\label{sec3}

The overall attenuation ($\alpha$) of optical modes in low-confinement waveguides arises from both intrinsic material properties and device geometry. Advances in fabrication have substantially reduced material absorption losses~\cite{Corato-Zanarella:24}, making geometry-dependent losses the primary contributors to total attenuation. These losses include (i) actuator absorption, (ii) scattering from interfacial roughness, (iii) radiation from waveguide bends, (iv) mode coupling loss at the fiber-to-waveguide interface, and (v) crosstalk between adjacent waveguides. Among these, absorption and scattering losses are particularly sensitive to waveguide dimensions (see Supplementary Note 6 and Supplementary Fig.~9), as variations in core and cladding size can significantly alter modal overlap with absorptive or rough regions. We combine FEM simulations with analytical loss models to estimate the propagation loss in our low-confinement platform, as described in Supplementary Note 7 and illustrated in Supplementary Fig.~9 and Fig.~10. Using the estimated photoelastic constants, we identify an optimal waveguide geometry by minimizing the voltage–loss product, as discussed in Supplementary Note 6 and Note 8 and in Supplementary Fig.~10.

\begin{figure}
\centering
\includegraphics[width=0.9\linewidth]{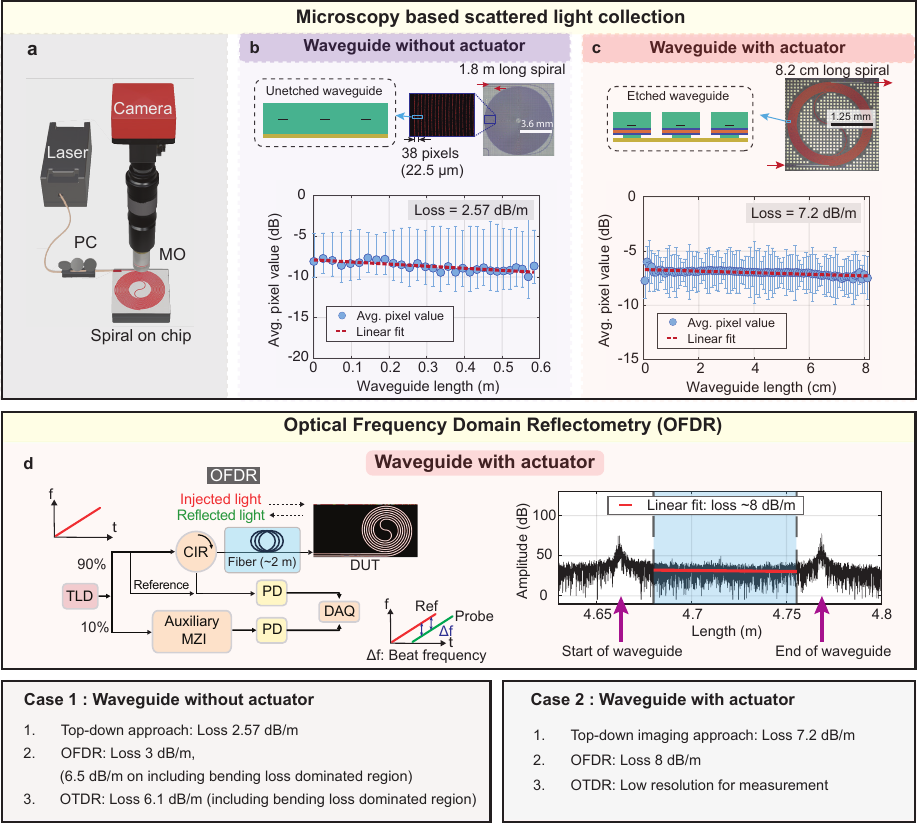}
\caption{\textbf{Waveguide-loss measurement using top-view imaging.}\\
\textbf{(a)} Schematic of the experimental setup for measuring waveguide loss using a top‑down micrograph technique. The setup includes a $780~\mathrm{nm}$ laser, polarization controller (PC), and a camera equipped with a $12\times$ zoom lens and a $5\times$ microscope objective (MO). \textbf{(b)} Propagation loss characterization of an unetched $1.8~\mathrm{m}$ silicon nitride Archimedean spiral waveguide (no actuator). Image analysis of a spiral section shows the average pixel intensity versus propagation distance; linear fitting yields a loss of $2.57~\mathrm{dB/m}$. \textbf{(c)} Propagation loss analysis of an etched $8.2~\mathrm{cm}$-long Archimedean spiral waveguide with an integrated bottom actuator, occupying a chip footprint of $6.25~\mathrm{mm^{2}}$ ($2.5~\mathrm{mm} \times 2.5~\mathrm{mm}$). Image processing of the complete spiral structure and pixel-intensity averaging along its length yield a propagation loss of $7.2~\mathrm{dB/m}$. \textbf{(d)} OFDR propagation loss measurement of an $8.2\,\mathrm{cm}$ waveguide with actuator. The tunable laser diode (TLD) output is divided into a reference arm and a measurement arm comprising a $2\,\mathrm{m}$ delay fiber, circulator (CIR), and device under test (DUT), with a fiber Mach--Zehnder interferometer (MZI) for frequency calibration. The OFDR trace reveals Fresnel reflection peaks at the input/output facets; a linear fit to the Rayleigh backscatter yields a one-way propagation loss of ${\sim}8~\mathrm{dB/m}$. The fit is taken over the blue-shaded region, which corresponds to the spiral waveguide section with the integrated actuator.} 
\label{Fig4}
\end{figure}

We characterized the propagation loss using a top-down imaging technique (Ref.~\cite{Jaber:24}, Fig.~~\ref{Fig4}a) in two distinct low-confinement waveguide geometries: (i) an unetched waveguide without an actuator (Fig.~~\ref{Fig4}b) and (ii) an etched waveguide with a bottom actuator (Fig.~~\ref{Fig4}c). In the unetched device, the dominant loss mechanism is scattering at the core–cladding interface due to fabrication imperfections. By contrast, the etched waveguide with an integrated actuator exhibits additional losses from actuator absorption as well as scattering at both the core–cladding and cladding–air interfaces. The top-down imaging technique (Fig.~~\ref{Fig4}a) collects light scattered perpendicular to the direction of optical mode propagation. This method is based on the principle that the local intensity of scattered light is directly proportional to the remaining guided power within the waveguide. For the unetched waveguide shown in  Fig.~~\ref{Fig4}b, we image a section of the 1.8-m-long Archimedean spiral, where adjacent waveguides in the image are separated by approximately $38$ pixels, corresponding to $2.4~\mathrm{cm}$ of waveguide length along the spiral. We apply a $20$-pixel-wide analysis window following the waveguide curvature to calculate the average pixel intensity for each waveguide segment. By plotting these intensity values against the propagation length of the optical mode along the spiral and performing a linear fit, we extract a propagation loss of $2.57~\mathrm{dB/m}$.
For the etched waveguide with actuator shown in (Fig.~~\ref{Fig4}c), we image the entire 8.2-cm-long spiral in a single field of view and average pixel intensities along its length. We divide the spiral into 100 sections and each 0.82-mm-long section corresponds to a data point in the (Fig.~~\ref{Fig4}c) plot. In each section, we apply a $15$-pixel averaging window that covers the waveguide. Linear fitting of these data yields a propagation loss of $7.2~\mathrm{dB/m}$ for the actuator-integrated waveguide, $\sim4.63~\mathrm{dB/m}$ higher than that of the unetched waveguide. Cross-sectional SEM images of the waveguide in Fig.~\ref{Fig2}a reveal rough etched sidewalls located $6$--$7~\mu\mathrm{m}$ from the waveguide core. Finite-element simulations indicate that direct absorption by the actuator top metal layer is small ($\sim0.05$ dB/m) at a metal--waveguide separation of $3~\mu\mathrm{m}$. We therefore attribute the excess loss primarily to scattering at the cladding–metal interface, caused by nanoscale metal protrusions that extend into the oxide cladding. Smaller contributions arise from metal absorption and scattering at the cladding–air interface, as analyzed in detail in Supplementary Note 9.

We further verify the propagation loss values using optical frequency-domain reflectometry (OFDR), which sweeps a tunable laser from $773.2$ to $781~\mathrm{nm}$, yielding a spatial resolution of approximately $27~\mu\mathrm{m}$. Because of this resolution, OFDR is capable of characterizing both the actuator-free and actuator-integrated waveguides, mirroring the device set used in the top-down imaging measurements. For the $1.8~\mathrm{m}$-long actuator-free waveguide, OFDR measures a loss of $\sim2.96~\mathrm{dB/m}$ in the large bend-radius sections, in close agreement with the $2.57~\mathrm{dB/m}$ obtained from top-down imaging. When the smaller-bend-radius sections close to the spiral center, where bending loss is high, are included in the fit, the extracted loss increases to approximately $6.5~\mathrm{dB/m}$. For the $8.2~\mathrm{cm}$-long actuator-integrated waveguide, OFDR yields a loss of ${\sim}8$--$8.3~\mathrm{dB/m}$, in close agreement with the 
$7.2~\mathrm{dB/m}$ obtained from top-down imaging, with experimental details and data provided in Fig.~\ref{Fig4}d and Supplementary Note~5.3. 
We additionally cross-validate the actuator-free result using optical 
time-domain reflectometry (OTDR) with a femtosecond source, which has a 
coarser spatial resolution of $2.25~\mathrm{mm}$. This resolution is 
insufficient to characterize the $8.2~\mathrm{cm}$-long actuator-integrated waveguide, restricting OTDR's use to the long, actuator-free spiral, where it yields a higher apparent loss of $6.1~\mathrm{dB/m}$, because the fitted slope includes contributions from the high bending loss dominated central region of the spiral. This value is consistent with the OFDR measurement when the same bend-dominated region is included, as discussed in Supplementary Note~5.3 and Supplementary Figs.~6--8. Across all measurements, the actuator-free waveguide shows a propagation loss of $\leq 3~\mathrm{dB/m}$, while the actuator-integrated waveguide exhibits approximately $7.2$--$8.3~\mathrm{dB/m}$, in consistent agreement across top-down imaging, OFDR, and OTDR, as discussed in Supplementary Note~5.

\section*{Discussion}
\label{sec4}

\begin{figure}
\centering
\includegraphics[width=\linewidth]{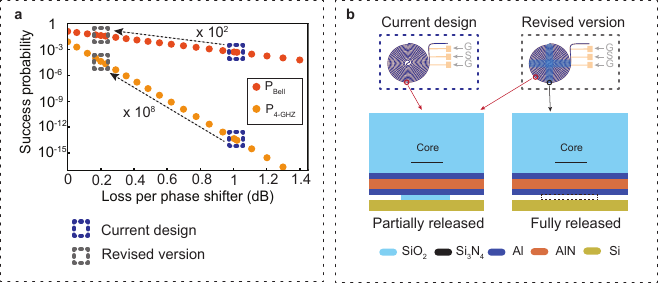}  
\caption{\textbf{Simulated performance scaling and proposed SPS geometry.}\\
\textbf{(a)} Success probability for generating heralded Bell states and 4-mode GHZ states within a lossy LPNP circuit. The arrows highlight the significant improvement in success probability achieved by reducing the loss per phase shifter from $1~\mathrm{dB}$ to $0.2~\mathrm{dB}$; this reduction yields probability enhancement factors of $\sim10^2$ for Bell states and $\sim10^8$ for 4-mode GHZ states (see Supplementary Note 4 for details). \textbf{(b)} Schematic of two SPS designs: the presently demonstrated device (partially released) and a proposed revised version (with fully released sections). The revised design aims to minimize the propagation loss per phase shifter in the LPNP circuit toward the low-loss regime ($0.2~\mathrm{dB}$) simulated in \textbf{(a)}.}
\label{Fig3a}
\end{figure}

Core thickness governs a fundamental trade-off in piezo-optomechanical phase shifters: a thicker core reduces the mode area and allows a smaller core--actuator separation, improving actuator responsivity (lower $V_{\pi} L$), but simultaneously increases scattering loss at the core–cladding interface due to the tighter mode confinement, raising the propagation loss $\alpha$. We use the voltage--loss product $\mathrm{VLP} = V_{\pi}L \times \alpha$ as a figure of merit to capture both effects in a single metric and enable direct comparison of platforms across this trade-off. Our $50~\mathrm{nm}$-core, low-confinement waveguides, fabricated via PECVD, demonstrate that low $\alpha$ can outweigh the penalty of a larger $V_{\pi} L$. Leveraging the propagation losses established above ($\leq 3~\mathrm{dB/m}$ without an actuator and $7.2$--$8.3~\mathrm{dB/m}$ with an actuator), we implement $8.7~\mathrm{cm}$-long phase shifters with $0.626~\mathrm{dB}$ propagation loss, achieving a voltage--length product of $\sim 2.8~\mathrm{V{\cdot}m}$ at $780~\mathrm{nm}$ and $\mathrm{VLP} \sim 20.2~\mathrm{V{\cdot}dB}$. This represents a clear improvement over a high-confinement platform that uses the same dominant phase-shifting mechanism as ours (strain-induced change in the effective refractive index) and reports $\mathrm{VLP} \approx 50.7~\mathrm{V}\cdot\mathrm{dB}$ with propagation losses of $35$--$100~\mathrm{dB/m}$~\cite{Dong22_01}. Our current $\mathrm{VLP}$ is comparable to that of another high-confinement piezo-optomechanical platform ($22~\mathrm{V}\cdot\mathrm{dB}$, reported in Ref.~\cite{Dong22_05}), which relies predominantly on strain-induced optical path length modulation rather than the photoelastic effect. However, our low-confinement waveguides exhibit propagation losses significantly lower than those reported in Ref.~\cite{Dong22_05} ($7.2~\mathrm{dB/m}$ in our platform vs.\ $\sim38~\mathrm{dB/m}$). Moreover, our platform can leverage strain-induced optical path length modulation (the dominant mechanism in Ref.~\cite{Dong22_05}) through selective release of SPS sections (Fig.~\ref{Fig3a}b and Supplementary Figs.~1a,b) to further reduce the $\mathrm{VLP}$. Simulations shown in Supplementary Figs.~1a,b indicate that fully releasing selected SPS sections could reduce the $\mathrm{VLP}$ by a factor of $5$--$10\times$ relative to the current devices, potentially lowering it to $\sim4~\mathrm{V}\cdot\mathrm{dB}$ or below. Importantly, this enhanced responsivity can be achieved even with larger core-to-actuator separations, which simultaneously reduce the evanescent field overlap with the metal electrodes, lowering the actuator-induced excess loss from the current $7.2~\mathrm{dB/m}$ toward the $\leq3~\mathrm{dB/m}$ baseline measured for actuator-free waveguides. This loss reduction alone would provide an additional $\sim2$--$3\times$ improvement in $\mathrm{VLP}$. Combining the responsivity gain from SPS release with the corresponding reduction in actuator-induced loss could therefore lower the $\mathrm{VLP}$ to approximately $0.7$--$2~\mathrm{V}\cdot\mathrm{dB}$.

We estimate the impact of reduced loss in LPNP circuits through an example protocol for the on-demand generation of heralded Bell and GHZ states using four-mode and eight-mode circuits~\cite{Clements16,bhatti2023GHZwithSMS,bhatti2024heraldingGHZ}. For this analysis, we consider propagation loss in the phase shifters as the sole loss mechanism, assuming all other components are lossless. The loss per phase shifter, $\gamma$, is equal to $\mathrm{VLP} / \mathrm{V}_{\mathrm{max}}$. At $\mathrm{V}_{\mathrm{max}}=20~\mathrm{V}$ (typical benchtop function generator maximum), our current low-confinement platform ($\mathrm{VLP}\sim20.2~\mathrm{V}\cdot\mathrm{dB}$) incurs $\sim1~\mathrm{dB}$ per phase shifter, comparable to the high-confinement platform of Ref.~\cite{Dong22_05}, whereas a future low-confinement platform (Fig.~\ref{Fig3a}b and Supplementary Fig.~1) with $\mathrm{VLP}<4~\mathrm{V}\cdot \mathrm{dB}$ would suffer a loss of $<0.2~\mathrm{dB}$ per phase shifter. Fig.~\ref{Fig3a}a illustrates how this difference translates into the probability of successfully heralding quantum states in a lossy LPNP circuit with ideal detectors. The results indicate dramatic improvements in quantum-state generation efficiency. For Bell-state preparation in a four-mode LPNP circuit, the success probability increases by roughly two orders of magnitude with the lower phase-shifter loss. The improvement is even more striking for heralded four-mode GHZ states, where the success probability increases by a factor of $>10^{8}$ (Supplementary Note 4 and Supplementary Fig.~5). This advantage becomes even more pronounced when compared against high-confinement platforms with $\sim2.5~\mathrm{dB}$ loss per phase shifter, where the gap in GHZ-state generation efficiency spans several additional orders of magnitude.

 This low-loss, scalable architecture forms the essential hardware backbone for applications such as quantum routers~\cite{Lee2022}, universal quantum photonic logic circuits~\cite{Taballione21_08, Carolan15}, on-chip quantum memories for heralded entanglement swapping~\cite{PhysRevResearch.5.033149}, tunable low-power optical true-time delay lines~\cite{Ji:19}, and post-selected heralded GHZ state generation~\cite{bhatti2024heraldingGHZ, bhatti2023GHZwithSMS}. Furthermore, the platform’s ultra-low propagation loss and anneal-free, foundry-compatible fabrication process enable seamless integration of single-photon sources, modulators, and electronic control, paving the way for complex, scalable, and fully integrated quantum photonic systems on a single chip.

\section*{Methods}

\subsection*{Devices design}

We characterized the SPS by integrating it into one arm of an unbalanced fiber Mach--Zehnder interferometer, as illustrated in Fig.~\ref{Fig2}c. The three fabricated SPS samples have waveguide lengths of $8.2~\mathrm{cm}$, $8.7~\mathrm{cm}$, and $9.8~\mathrm{cm}$, with minimum bend radii of $500~\mu\mathrm{m}$, $300~\mu\mathrm{m}$, and $150~\mu\mathrm{m}$, respectively. These designs occupy chip footprints of $6.25~\mathrm{mm}^2$ ($2.5~\mathrm{mm}\times2.5~\mathrm{mm}$), $4.84~\mathrm{mm}^2$ ($2.2~\mathrm{mm}\times2.2~\mathrm{mm}$), and $3.80~\mathrm{mm}^2$ ($1.95~\mathrm{mm}\times1.95~\mathrm{mm}$). Adjacent waveguides in each spiral are separated by an inter-waveguide gap of approximately $22.5~\mu\mathrm{m}$ and a released trench of about $3~\mu\mathrm{m}$ for the undercut.

The footprint of the Archimedean spiral is governed by two geometric parameters: the minimum bend radius and the inter-waveguide pitch. Reducing the bend radius shrinks the innermost fold of the spiral and thus the overall device area, but increases bending loss; our FEM simulations show that the bending loss rises from $\sim$$10^{-5}~\mathrm{dB/m}$ at a $500~\mu\mathrm{m}$ bend radius to $\sim$$10^{-1}~\mathrm{dB/m}$ at $150~\mu\mathrm{m}$, setting a practical lower bound on compactness. Similarly, reducing the waveguide pitch allows more turns within the same area but increases inter-waveguide crosstalk; our simulations confirm that crosstalk between two 1 mm long waveguides becomes negligible beyond $\sim$$10~\mu\mathrm{m}$ spacing, and our current pitch of $\sim$$22.5~\mu\mathrm{m}$ provides a conservative margin.

\subsection*{Devices measurement}

To characterize the devices, we drive the SPS using a Keysight 33500B function generator as a voltage source in combination with a Thorlabs BPA100 benchtop high-voltage piezo amplifier, which provides a noise bandwidth from $20~\mathrm{Hz}$ to $100~\mathrm{kHz}$. We apply a $200~\mathrm{Hz}$ ramp modulation signal to the SPS (Fig.~\ref{Fig2}c), sweeping the voltage up to $120~\mathrm{V}$. We perform S-parameter measurements using a Keysight P9374A vector network analyzer. We measure the holding power with a Keithley 2450 source-measure unit, as shown in Supplementary Fig.~4b.

\subsection*{Top-down imaging measurement}

Fig.~\ref{Fig4}a shows the experimental setup used to measure waveguide loss with a top-down imaging technique. We capture images using an Allied Vision 1800 U-120c camera (1280 × 960 pixels) equipped with a Navitar $12\times$ UltraZoom Video Microscopy System and a Mitutoyo $5\times$ microscope objective (MO). Because the low-confinement waveguides exhibit ultra-low loss and therefore weak light scattering, we adjust the imaging parameters during acquisition to improve the signal-to-noise ratio. For the unetched waveguide, where we image a section of the $1.8~\mathrm{m}$-long spiral, we use a gamma correction of 0.5, a gain of $35~\mathrm{dB}$, and an exposure time of $8.03~\mathrm{ms}$. For the etched waveguide with an actuator, we set the gamma correction to 0.5, the gain to $38.3~\mathrm{dB}$, and the exposure time to $7.3~\mathrm{ms}$.

\section*{Acknowledgments}
This material is based upon work supported by the U.S. Department of Energy, Office of Science, National Quantum Information Science Research Centers, Quantum Systems Accelerator. 
\section*{Author contributions:} 
N.T.O. and M.E. conceived the work. M.M. and G.M.T. performed simulations to inform device design, and M.M. designed the device layouts for fabrication. D.D. and A.L. coordinated device fabrication. M.M., G.C. and K.K. contributed to the experiments, and M.M., G.C., W.H., M.G., N.T.O. and M.E. contributed to data analysis and simulations. M.M. drafted the manuscript. M.M., G.C., W.H., G.M.T., N.T.O. and M.E. contributed to manuscript revision. All authors discussed the results and implications.
\section*{Competing interests:} 
The authors declare no competing interests.
\section*{Data and materials availability:} 
All data supporting the findings of this study are available within the Article, Supplementary Information, and Source Data file. Additional data and code are available from the corresponding author upon request.

\bibliography{bibliography}

\begin{thebibliography}{99}

\bibitem{SIMorin:21a}
T.~J.~Morin, L.~Chang, W.~Jin, C.~Li, J.~Guo, H.~Park, M.~A.~Tran, T.~Komljenovic, and J.~E.~Bowers,
``CMOS-foundry-based blue and violet photonics,''
\textit{Optica}, vol.~8, no.~5, pp.~755--756, May 2021.
\href{https://doi.org/10.1364/OPTICA.426065}{doi:10.1364/OPTICA.426065}.

\bibitem{SIChauhan:22}
N.~Chauhan, J.~Wang, D.~Bose, K.~Liu, R.~L.~Compton, C.~Fertig, C.~W.~Hoyt, and D.~J.~Blumenthal,
``Ultra-low loss visible light waveguides for integrated atomic, molecular, and quantum photonics,''
\textit{Opt. Express}, vol.~30, no.~5, pp.~6960--6969, Feb. 2022.
\href{https://doi.org/10.1364/OE.448938}{doi:10.1364/OE.448938}.

\bibitem{SInick26_Blumenthal}
N.~Montifiore \textit{et al.},
``Blue to near-IR integrated PZT silicon nitride modulators for quantum and atomic applications,''
arXiv:2601.15695, 2026.
\url{https://arxiv.org/abs/2601.15695}


\bibitem{SILiu2024a}
K.~Liu, N.~Chauhan, M.~Song, M.~W.~Harrington, K.~D.~Nelson, and D.~J.~Blumenthal,
``Tunable broadband two-point-coupled ultra-high-Q visible and near-infrared photonic integrated resonators,''
\textit{Photonics Res.}, vol.~12, no.~9, pp.~1890--1899, 2024.
\href{https://doi.org/10.1364/PRJ.528398}{doi:10.1364/PRJ.528398}.

\bibitem{SIJaber2024}
N.~Jaber, S.~Madaras, A.~Starbuck, A.~Pomerene, C.~Dallo, D.~C.~Trotter, M.~Gehl, and N.~Otterstrom,
``Non-resonant Bragg scattering four-wave mixing at near-visible wavelengths in low-confinement silicon nitride waveguides,''
\textit{Opt. Lett.}, vol.~49, pp.~3146--3149, 2024.

\bibitem{SIProkoshin2024}
A.~Prokoshin, M.~Gehl, S.~Madaras, W.~W.~Chow, and Y.~Wan,
``Ultra-narrow-linewidth hybrid-integrated self-injection locked laser at 780~nm,''
\textit{Optica}, vol.~11, pp.~1024--1029, 2024.

\bibitem{SIIsichenko2024a}
A.~Isichenko, N.~Chauhan, K.~Liu, M.~W.~Harrington, and D.~J.~Blumenthal,
``Sub-Hz fundamental, sub-kHz integral linewidth self-injection locked 780\,nm hybrid integrated laser,''
\textit{Sci. Rep.}, vol.~14, p.~27015, 2024.
\href{https://doi.org/10.1038/s41598-024-76699-x}{doi:10.1038/s41598-024-76699-x}.

\bibitem{Majumdar2025LC}
A.~Majumdar \textit{et al.},
``Near-visible low power tuning of nematic-liquid crystal integrated silicon nitride ring,''
arXiv:2503.21012, 2025.
\url{https://arxiv.org/abs/2503.21012}.
 
\bibitem{SINatCom:22}
A.~Chanana, H.~Larocque, R.~Moreira \textit{et al.},
``Ultra-low loss quantum photonic circuits integrated with single quantum emitters,''
\textit{Nat. Commun.}, vol.~13, p.~7693, Dec. 2022.
\href{https://doi.org/10.1038/s41467-022-35332-z}{doi:10.1038/s41467-022-35332-z}.

\bibitem{Bose:24}
D.~Bose, M.~W.~Harrington, A.~Isichenko, K.~Liu, J.~Wang, Z.~L.~Newman, and D.~J.~Blumenthal,
``Anneal-free ultra-low loss silicon nitride integrated photonics,''
\textit{Light Sci. Appl.}, vol.~13, p.~156, 2024.
\href{https://doi.org/10.1038/s41377-024-01503-4}{doi:10.1038/s41377-024-01503-4}.


\bibitem{Dullo2026}
F.~T.~Dullo \textit{et al.},
``Piezoelectric MEMS phase modulator for silicon nitride platform in the visible spectrum,''
arXiv:2602.16577, 2026.
\url{https://arxiv.org/abs/2602.16577}.

\bibitem{mohanty2020reconfigurable}
A. Mohanty, Q. Li, M. A. Tadayon, S. P. Roberts, G. R. Bhatt, E. Shim,
X. Ji, J. Cardenas, S. A. Miller, A. Kepecs, \textit{et al.},
``Reconfigurable nanophotonic silicon probes for sub-millisecond deep-brain optical stimulation,''
\textit{Nature Biomedical Engineering}, vol. 4, no. 2, pp. 223--231, 2020.

\bibitem{SILiang2021a}
G.~Liang, H.~Huang, A.~Mohanty \textit{et al.},
``Robust, efficient, micrometre-scale phase modulators at visible wavelengths,''
\textit{Nat. Photon.}, vol.~15, pp.~908--913, 2021.
\href{https://doi.org/10.1038/s41566-021-00891-y}{doi:10.1038/s41566-021-00891-y}.

\bibitem{grimpe2026benchmarking}
C.-F. Grimpe, A. L{\"u}{\ss}mann-Sorokina, G. Du, P. Sah, S. Sauer,
E. Jordan, R. Thomas, P. Gehrmann, M. Lipkin, S. Suckow, \textit{et al.},
``Benchmarking dual-polarization silicon nitride photonic integrated circuits for trapped-ion quantum technologies,''
\textit{arXiv preprint arXiv:2603.22578}, 2026.




\bibitem{Sinclair:23}
G.~Sinclair, K.~Dideriksen, and P.~Lodahl,
``SiN foundry platform for high performance visible light integrated photonics,''
\textit{Opt. Mater. Express}, vol.~13, no.~2, pp.~458--468, 2023.

\bibitem{CorZan:24}
M.~Corato-Zanarella, X.~Ji, A.~Mohanty, and M.~Lipson,
``Absorption and scattering limits of silicon nitride integrated photonics in the visible spectrum,''
\textit{Opt. Express}, vol.~32, no.~4, pp.~5718--5728, 2024.
\href{https://doi.org/10.1364/OE.505892}{doi:10.1364/OE.505892}.

\bibitem{SISacher:19b}
W.~D.~Sacher, X.~Luo, Y.~Yang, F.-D.~Chen, T.~Lordello, J.~C.~C.~Mak \textit{et al.},
``Visible-light silicon nitride waveguide devices and implantable neurophotonic probes on thinned 200~mm silicon wafers,''
\textit{Opt. Express}, vol.~27, no.~26, pp.~37400--37418, Dec. 2019.
\href{https://doi.org/10.1364/OE.27.037400}{doi:10.1364/OE.27.037400}.


\bibitem{Muller:16}
C.~Muller \textit{et al.},
``Multisite silicon neural probes with integrated silicon nitride waveguides and gratings for optogenetic applications,''
\textit{Sci. Rep.}, vol.~6, p.~22693, 2016.
\href{https://doi.org/10.1038/srep22693}{doi:10.1038/srep22693}.

\bibitem{SIGovdeli:25}
A.~Govdeli \textit{et al.},
``Integrated photonic MEMS switch for visible light,''
\textit{Opt. Express}, vol.~33, no.~1, p.~650, 2025.

\bibitem{SIWangTO:24}
J.~Wang \textit{et al.},
``Submilliwatt silicon nitride thermo-optic modulator operating at 532\,nm,''
\textit{Photonics}, vol.~11, no.~3, p.~213, 2024.
\href{https://doi.org/10.3390/photonics11030213}{doi:10.3390/photonics11030213}.


\bibitem{Notaros2022LC}
M. Notaros, T. Dyer, M. Raval, C. Baiocco, J. Notaros, and M. R. Watts,
``Integrated visible-light liquid-crystal-based phase modulators,''
\textit{Optics Express}, vol. 30, no. 8, pp. 13790--13801, Apr. 2022,
doi: 10.1364/OE.454494.


\bibitem{SIHosseini:09}
E.~S.~Hosseini, S.~Yegnanarayanan, A.~H.~Atabaki, M.~Soltani, and A.~Adibi,
``High quality planar silicon nitride microdisk resonators for integrated photonics in the visible wavelength range,''
\textit{Opt. Express}, vol.~17, no.~17, pp.~14543--14551, Aug. 2009.
\href{https://doi.org/10.1364/OE.17.014543}{doi:10.1364/OE.17.014543}.



\bibitem{SIDoolin:15}
C.~Doolin, P.~Doolin, B.~C.~Lewis, and J.~P.~Davis,
``Refractometric sensing of Li salt with visible-light Si$_3$N$_4$ microdisk resonators,''
\textit{Appl. Phys. Lett.}, vol.~106, no.~8, p.~081104, Feb. 2015.
\href{https://doi.org/10.1063/1.4913618}{doi:10.1063/1.4913618}.

\bibitem{SISinclair:20}
M.~Sinclair, K.~Gallacher, M.~Sorel, J.~C.~Bayley, E.~McBrearty, R.~W.~Millar, S.~Hild, and D.~J.~Paul,
``1.4 million Q factor Si$_3$N$_4$ micro-ring resonator at 780~nm wavelength for chip-scale atomic systems,''
\textit{Opt. Express}, vol.~28, no.~3, pp.~4010--4020, Feb. 2020.
\href{https://doi.org/10.1364/OE.381224}{doi:10.1364/OE.381224}.

\bibitem{SIZhao:20}
Y.~Zhao, X.~Ji, B.~Y.~Kim \textit{et al.},
``Visible nonlinear photonics via high-order-mode dispersion engineering,''
\textit{Optica}, vol.~7, no.~2, pp.~135--141, Feb. 2020.
\href{https://doi.org/10.1364/OPTICA.7.000135}{doi:10.1364/OPTICA.7.000135}.

\bibitem{Buzaverov:23}
K.~A.~Buzaverov \textit{et al.},
``Low-loss silicon nitride photonic ICs for near-infrared wavelength bandwidth,''
\textit{Opt. Express}, vol.~31, no.~10, pp.~16227--16242, 2023.
\href{https://doi.org/10.1364/OE.477458}{doi:10.1364/OE.477458}.

\bibitem{3DSiNAlN:25}
J.~Zhang \textit{et al.},
``3D heterogeneous integration of silicon nitride and aluminum nitride photonic integrated circuits on sapphire,''
\textit{ACS Photonics}, vol.~12, pp.~2058--2067, 2025.
\href{https://doi.org/10.1021/acsphotonics.4c02556}{doi:10.1021/acsphotonics.4c02556}.

\bibitem{SIYong:22}
Z. Yong, H. Chen, X. Luo, A. Govdeli, H. Chua, S. S. Azadeh,
A. Stalmashonak, G.-Q. Lo, J. K. S. Poon, and W. D. Sacher,
``Power-efficient silicon nitride thermo-optic phase shifters for visible light,''
\textit{Optics Express}, vol. 30, pp. 7225--7237, 2022.

\bibitem{mcnulty2025overcoming}
K. J. McNulty, S. Chaitanya, S. Sanyal, A. Gil-Molina,
M. Corato-Zanarella, Y. Okawachi, A. L. Gaeta, and M. Lipson,
``Overcoming stress limitations in SiN nonlinear photonics via a bilayer waveguide,''
\textit{Nanophotonics}, vol. 14, no. 23, pp. 3921--3926, 2025.

\bibitem{SILe:26}
T.-L.~Le \textit{et al.},
``Hybrid electrostatic-piezo MEMS photonic integrated modulators,''
\textit{APL Photonics}, vol.~11, p.~031301, 2026.
\href{https://doi.org/10.1063/5.0316551}{doi:10.1063/5.0316551}.


\bibitem{SISubramanian:13}
A.~Z.~Subramanian, P.~Neutens, A.~Dhakal \textit{et al.},
``Low-loss singlemode PECVD silicon nitride photonic wire waveguides for 532--900~nm wavelength window fabricated within a CMOS pilot line,''
\textit{IEEE Photon. J.}, vol.~5, no.~6, p.~2202809, 2013.
\href{https://doi.org/10.1109/JPHOT.2013.2292698}{doi:10.1109/JPHOT.2013.2292698}.

\bibitem{SIRomeroGarcia:13}
S.~Romero-García, F.~Merget, F.~Zhong, H.~Finkelstein, and J.~Witzens,
``Silicon nitride CMOS-compatible platform for integrated photonics applications at visible wavelengths,''
\textit{Opt. Express}, vol.~21, no.~12, pp.~14036--14046, Jun. 2013.
\href{https://doi.org/10.1364/OE.21.014036}{doi:10.1364/OE.21.014036}.

\bibitem{SIDong2022}
M.~Dong, D.~Heim, A.~Witte \textit{et al.},
``Piezo-optomechanical cantilever modulators for VLSI visible photonics,''
\textit{APL Photonics}, vol.~7, no.~5, p.~051304, May 2022.
\href{https://doi.org/10.1063/5.0088424}{doi:10.1063/5.0088424}.

\bibitem{SIDong:22}
M.~Dong, G.~Clark, A.~J.~Leenheer \textit{et al.},
``High-speed programmable photonic circuits in a cryogenically compatible, visible--near-infrared 200~mm CMOS architecture,''
\textit{Nat. Photon.}, vol.~16, no.~1, pp.~59--65, Jan. 2022.
\href{https://doi.org/10.1038/s41566-021-00903-x}{doi:10.1038/s41566-021-00903-x}.

\bibitem{SIBhatti:25}
D.~Bhatti and S.~Barz,
``Heralding higher-dimensional Bell and Greenberger--Horne--Zeilinger states using multiport splitters,''
\textit{New J. Phys.}, vol.~27, no.~3, p.~033006, 2025.

\bibitem{SIClements:16}
W.~R.~Clements, P.~C.~Humphreys, B.~J.~Metcalf, W.~S.~Kolthammer, and I.~A.~Walmsley,
``Optimal design for universal multiport interferometers,''
\textit{Optica}, vol.~3, no.~12, pp.~1460--1465, Dec. 2016.
\href{https://doi.org/10.1364/OPTICA.3.001460}{doi:10.1364/OPTICA.3.001460}.

\bibitem{SIBarwicz:05}
T.~Barwicz and H.~A.~Haus,
``Three-dimensional analysis of scattering losses due to sidewall roughness in microphotonic waveguides,''
\textit{J. Lightw. Technol.}, vol.~23, no.~9, pp.~2719--2732, 2005.
\href{https://doi.org/10.1109/JLT.2005.850816}{doi:10.1109/JLT.2005.850816}.

\bibitem{SIBauters:13}
J.~F.~Bauters,
``Ultra-low loss waveguides with application to photonic integrated circuits,''
Ph.D. dissertation, Univ. of California, Santa Barbara, 2013.

\bibitem{SIBauters:11}
J.~F.~Bauters, M.~J.~R.~Heck, D.~John, D.~Dai, M.-C.~Tien, J.~S.~Barton, A.~Leinse, R.~G.~Heideman, D.~J.~Blumenthal, and J.~E.~Bowers,
``Ultra-low-loss high-aspect-ratio Si$_3$N$_4$ waveguides,''
\textit{Opt. Express}, vol.~19, no.~4, pp.~3163--3174, Feb. 2011.
\href{https://doi.org/10.1364/OE.19.003163}{doi:10.1364/OE.19.003163}.

\bibitem{SIKuznetsov:83}
M.~Kuznetsov and H.~Haus,
``Radiation loss in dielectric waveguide structures by the volume current method,''
\textit{IEEE J. Quantum Electron.}, vol.~19, no.~10, pp.~1505--1514, 1983.
\href{https://doi.org/10.1109/JQE.1983.1071758}{doi:10.1109/JQE.1983.1071758}.

\bibitem{SIChen2019}
H.~Chen, H.~Fu, J.~Zhou, X.~Huang, T.-H.~Yang, K.~Fu, C.~Yang, J.~A.~Montes, and Y.~Zhao,
``Study of crystalline defect induced optical scattering loss inside photonic waveguides in UV-visible spectral wavelengths using volume current method,''
\textit{Opt. Express}, vol.~27, no.~12, pp.~17262--17273, Jun. 2019.
\href{https://doi.org/10.1364/OE.27.017262}{doi:10.1364/OE.27.017262}.

\bibitem{SIGyger2020a}
F.~Gyger, J.~Liu, F.~Yang \textit{et al.},
``Observation of stimulated Brillouin scattering in silicon nitride integrated waveguides,''
\textit{Phys. Rev. Lett.}, vol.~124, no.~1, p.~013902, Jan. 2020.
\href{https://doi.org/10.1103/PhysRevLett.124.013902}{doi:10.1103/PhysRevLett.124.013902}.

\bibitem{SIKhalatpour2025}
A.~Khalatpour, L.~Qi, M.~M.~Fejer, and A.~Safavi-Naeini,
``Roughness-limited performance in ultra-low-loss lithium niobate cavities,''
arXiv:2505.01913, 2025.
\url{https://arxiv.org/abs/2505.01913}

\bibitem{SITian:24}
H.~Tian, J.~Liu, A.~Attanasio, A.~Siddharth, T.~Bl\'{e}sin, R.~N.~Wang, A.~Voloshin, G.~Lihachev, J.~Riemensberger, S.~E.~Kenning, Y.~Tian, T.~H.~Chang, A.~Bancora, V.~Snigirev, V.~Shadymov, T.~J.~Kippenberg, and S.~A.~Bhave,
``Piezoelectric actuation for integrated photonics,''
\textit{Adv. Opt. Photon.}, vol.~16, no.~4, pp.~749--867, 2024.

\bibitem{SILacey:90}
J.~Lacey and F.~Payne,
``Radiation loss from planar waveguides with random wall imperfections,''
\textit{IEE Proc. J (Optoelectron.)}, vol.~137, pp.~282--289, 1990.













\bibitem{Renaud2023}
D.~Renaud \textit{et al.},
``Sub-1-volt and high-bandwidth visible to near-infrared electro-optic modulators,''
\textit{Nat.\ Commun.}, vol.~14, art.~1496, 2023.
\href{https://doi.org/10.1038/s41467-023-36870-w}{doi:10.1038/s41467-023-36870-w}.

\bibitem{Celik2022}
O.~T.~Celik \textit{et al.},
``High-bandwidth CMOS-voltage-level electro-optic modulation of 780\,nm light in thin-film lithium niobate,''
\textit{Opt.\ Express}, vol.~30, no.~13, pp.~23177--23186, 2022.
\href{https://doi.org/10.1364/OE.460119}{doi:10.1364/OE.460119}.

\bibitem{CNF2023}
Cornell NanoScale Facility,
``Full-spectrum visible electro-optic modulator,''
\textit{CNF Research Accomplishments}, pp.~102--103, 2023.
\url{https://www.cnf.cornell.edu/sites/default/files/2023-RA/2023cnfRA_PG102.pdf}.

\bibitem{Li:22}
C. Li, B. Chen, Z. Ruan, H. Wu, Y. Zhou, J. Liu, P. Chen, K. Chen,
C. Guo, and L. Liu,
``High modulation efficiency and large bandwidth thin-film lithium niobate modulator for visible light,''
\textit{Optics Express}, vol. 30, no. 20, pp. 36394--36402, Sep. 2022,
doi: 10.1364/OE.469065.

\bibitem{Desiatov:19}
B. Desiatov, A. Shams-Ansari, M. Zhang, C. Wang, and M. Lon\v{c}ar,
``Ultra-low-loss integrated visible photonics using thin-film lithium niobate,''
\textit{Optica}, vol. 6, no. 3, pp. 380--384, Mar. 2019,
doi: 10.1364/OPTICA.6.000380.

\bibitem{Powell2025LT}
K.~Powell \textit{et al.},
``A sub-volt near-IR lithium tantalate electro-optic modulator,''
\textit{APL Photon.}, vol.~10, art.~091303, 2025.
\href{https://doi.org/10.1063/5.0278724}{doi:10.1063/5.0278724}.

\bibitem{Pernice2012}
W.~H.~P.~Pernice \textit{et al.},
``Low-loss, silicon integrated, aluminum nitride photonic circuits and their use for electro-optic signal processing,''
\textit{Nano Lett.}, vol.~12, no.~6, pp.~3416--3421, 2012.
\href{https://doi.org/10.1021/nl3011885}{doi:10.1021/nl3011885}.


\bibitem{Zhu2020AlN}
D.~Zhu \textit{et al.},
``Aluminum nitride ultralow-loss waveguides and push-pull electro-optic modulators,''
in \textit{Proc.\ Opt.\ Fiber Commun.\ Conf.\ (OFC)}, paper T3H.5, 2020.
\href{https://doi.org/10.1364/OFC.2020.T3H.5}{doi:10.1364/OFC.2020.T3H.5}.

\bibitem{Wooten2000}
E.~L.~Wooten \textit{et al.},
``A review of lithium niobate modulators for fiber-optic communications systems,''
\textit{IEEE J.\ Sel.\ Top.\ Quantum Electron.}, vol.~6, no.~1, pp.~69--82, 2000.
\href{https://doi.org/10.1109/2944.826874}{doi:10.1109/2944.826874}.

\bibitem{Kamada:22}
S. Kamada, R. Ueda, C. Yamada, K. Tanaka, T. Yamada, and A. Otomo,
``Superiorly low half-wave voltage electro-optic polymer modulator for visible photonics,''
\textit{Optics Express}, vol. 30, no. 11, pp. 19771--19780, May 2022,
doi: 10.1364/OE.456271.



\bibitem{Gyger:21}
S. Gyger, J. Zichi, L. Schweickert, \textit{et al.},
``Reconfigurable photonics with on-chip single-photon detectors,''
\textit{Nature Communications}, vol. 12, article no. 1408, 2021, doi: 10.1038/s41467-021-21624-3.



\bibitem{liang2021robust}
G. Liang, H. Huang, A. Mohanty, M. C. Shin, X. Ji, M. J. Carter,
S. Shrestha, M. Lipson, and N. Yu,
``Robust, efficient, micrometre-scale phase modulators at visible wavelengths,''
\textit{Nature Photonics}, vol. 15, no. 12, pp. 908--913, 2021.

\bibitem{Freedman2025}
J.~M.~Freedman \textit{et al.},
``Gigahertz-frequency acousto-optic phase modulation of visible light in a CMOS-fabricated photonic circuit,''
\textit{Nat.\ Commun.}, vol.~16, art.~10959, 2025.
\href{https://doi.org/10.1038/s41467-025-65937-z}{doi:10.1038/s41467-025-65937-z}.

\bibitem{SIStanfield19_09}
P.~R.~Stanfield, A.~J.~Leenheer, C.~P.~Michael, R.~Sims, and M.~Eichenfield, 
``CMOS-compatible, piezo-optomechanically tunable photonics for visible wavelengths and cryogenic temperatures,''
\textit{Opt. Express}, vol.~27, no.~20, pp.~28588--28605, Sep. 2019.  
Available: \url{https://opg.optica.org/oe/abstract.cfm?URI=oe-27-20-28588}









\end{thebibliography}

\newpage
\clearpage

\let\origbibliography\bibliography
\renewcommand{\bibliography}[1]{}  

\setcounter{page}{1}
\pagenumbering{arabic}
\setcounter{section}{0}

\setcounter{figure}{0}
\renewcommand{\theHfigure}{supp.\arabic{figure}}


\definecolor{lpcvdcol}{HTML}{F3E2D4}   
\definecolor{pecvdcol}{HTML}{E3EDF7}   
\definecolor{thiswork}{HTML}{E3EEDC}   

\begin{center}

{\Large \textbf{Supplementary Information}}

\vspace{1.5em}

{\large \textbf{Ultra-low loss piezo-optomechanical low-confinement silicon nitride platform for visible wavelength quantum photonic circuits}}

\vspace{1.5em}

Mayank Mishra$^{1}$,
Gwangho Choi$^{1}$,
Wenhua He$^{1}$,
Gina M. Talcott$^{2}$,
Katherine Kearney$^{2}$,
Michael Gehl$^{2}$,
Andrew Leenheer$^{2}$,
Daniel Dominguez$^{2}$,
Nils T. Otterstrom$^{2*}$,
Matt Eichenfield$^{1,2,3*}$

\vspace{1em}

$^{1}$ James C. Wyant College of Optical Sciences, University of Arizona, Tucson, Arizona 85721, USA  

$^{2}$ Microsystems Engineering, Science, and Applications, Sandia National Laboratories, Albuquerque, New Mexico 87185, USA  

$^{3}$ Electrical, Computer and Energy Engineering, University of Colorado, Boulder, Colorado 80309, USA  

\vspace{1em}

*Correspondence to: ntotter@sandia.gov; Matt.Eichenfield@colorado.edu

\end{center}

\hypersetup{linktoc=none}
{\linespread{1.15}\selectfont
\tableofcontents
}

\newpage


\section{Visible Wavelength $\mathrm{Si}_3\mathrm{N}_4$ Platform Comparison}
\label{Table}

\setlength{\arrayrulewidth}{1.5pt}

\begin{table}[ht!]
\centering
\resizebox{\textwidth}{!}{\normalsize
\begin{tabular}{|p{2.5cm}|p{2.2cm}|p{2.2cm}|p{3.0cm}|p{2cm}|p{2.5cm}|p{3.5cm}|p{2cm}|}
\hline
\textbf{Material} & \textbf{Confinement} & \textbf{CMOS-BEOL Compatible} & \textbf{Fabrication / Mechanism} & \textbf{Wavelength (nm)} & \textbf{Propagation Loss (dB/m)} & \textbf{Voltage--loss product (V$\cdot$dB)} & \textbf{References}\\
\hline

\rowcolor{lpcvdcol}
Si$_3$N$_4$ & Low & No & LPCVD & 405, 450 & 93, 22 & N/A (passive) & \cite{SIMorin:21a} \\ \hline
\rowcolor{lpcvdcol}
Si$_3$N$_4$ & Low & No & LPCVD & 450, 461, 674, 698, 802 & 8, 9, 1, 3, 2 & N/A (passive) & \cite{SIChauhan:22} \\ \hline
\rowcolor{lpcvdcol}
Si$_3$N$_4$ & Low & No & LPCVD + PZT (stress-optic) & 493, 532, 780 & 24, 24, 27 & 2.34, 6.71, 2.39 & \cite{SInick26_Blumenthal} \\ \hline
\rowcolor{lpcvdcol}
Si$_3$N$_4$ & Low & No & LPCVD & 698, 780 & 0.83, 1.2 & N/A (passive/ active) & \cite{SILiu2024a} \\ \hline
\rowcolor{lpcvdcol}
Si$_3$N$_4$ & Low & No & LPCVD & 780 & 1.75 & N/A (passive) & \cite{SIJaber2024} \\ \hline
\rowcolor{lpcvdcol}
Si$_3$N$_4$ & Low & No & LPCVD & 780 & 1.5 & N/A (passive) & \cite{SIProkoshin2024} \\ \hline
\rowcolor{lpcvdcol}
Si$_3$N$_4$ & Low & No & LPCVD & 780 & 0.57 & N/A (passive/ active) & \cite{SIIsichenko2024a} \\ \hline
\rowcolor{lpcvdcol}
Si$_3$N$_4$ + Liquid crystal& Low & No & LPCVD & 780 & N/A & N/A (0.014\,V$\cdot$cm; 2.1\,nW) & \cite{Majumdar2025LC} \\ \hline
\rowcolor{lpcvdcol}
Si$_3$N$_4$ & Low & No & LPCVD & 930 & 1 & N/A (passive) & \cite{SINatCom:22} \\ \hline

\rowcolor{pecvdcol}
Si$_3$N$_4$ & Low & Yes & PECVD (anneal-free) & 780 & 1.8 & N/A (passive) & \cite{Bose:24} \\ \hline
\rowcolor{thiswork}
\textbf{Si$_3$N$_4$} & \textbf{Low} & \textbf{Yes} & \textbf{PECVD (anneal-free)} & \textbf{780} & \textbf{2.57} & \textbf{N/A (passive)} & \textbf{This work} \\ \hline
\rowcolor{thiswork}
\textbf{Si$_3$N$_4$} & \textbf{Low} & \textbf{Yes} & \textbf{PECVD (anneal-free)} & \textbf{780} & \textbf{7.2} & \textbf{20.2} & \textbf{This work} \\ \hline


\rowcolor{lpcvdcol}
Si$_3$N$_4$ & High & No & LPCVD + PZT-MEMS & 635 & 75 & 1.7 & \cite{Dullo2026} \\ \hline

\rowcolor{lpcvdcol}
Si$_3$N$_4$ & High & No & LPCVD & 473 & $<$300 & N/A (T-O, $P_\pi$ = 20--30\,mW) & \cite{mohanty2020reconfigurable} \\ \hline

\rowcolor{lpcvdcol}
Si$_3$N$_4$ & High & No & LPCVD & 488--530 & 260--960 & N/A (T-O, $P_\pi$= 0.85–1.8\,mW) & \cite{SILiang2021a} \\ \hline

\rowcolor{lpcvdcol}
Si$_3$N$_4$ & High & No & LPCVD & 760 & 291/205 (TE/TM) & N/A (passive) & \cite{grimpe2026benchmarking} \\ \hline


\rowcolor{lpcvdcol}
Si$_3$N$_4$ & High & No & LPCVD & 450--630, 630--850 & 100--300, ${<}100$ & N/A (passive) & \cite{Sinclair:23} \\ \hline
\rowcolor{lpcvdcol}
Si$_3$N$_4$ & High & No & LPCVD & 461, 532, 633, 730 & 4, 3, 2, 1 & N/A (passive) & \cite{CorZan:24} \\ \hline
\rowcolor{lpcvdcol}
Si$_3$N$_4$ & High & No & LPCVD & 466--550, 552--648, 602--648 & ${\leq}280$, ${\leq}190$, 50--140 & N/A (passive) & \cite{SISacher:19b} \\ \hline
\rowcolor{lpcvdcol}
Si$_3$N$_4$ & High & No & LPCVD & 473 & ${<}300$ & N/A (passive) & \cite{Muller:16} \\ \hline
\rowcolor{lpcvdcol}
Si$_3$N$_4$ + SOI & High & No & LPCVD & 540 & 250-350 & N/A (0.5\,nW static) & \cite{SIGovdeli:25} \\ \hline
\rowcolor{lpcvdcol}
Si$_3$N$_4$ & High & No & LPCVD & 532 & $\sim$300 & N/A (T-O, $P_\pi{=}0.63$\,mW) & \cite{SIWangTO:24} \\ \hline
\rowcolor{lpcvdcol}
Si$_3$N$_4$ + liquid crystal & High & No & LPCVD & 488 & N/A & N/A (0.006\,V$\cdot$cm) & \cite{Notaros2022LC} \\ \hline
\rowcolor{lpcvdcol}
Si$_3$N$_4$ & High & No & LPCVD & 654 & 25 & N/A (passive) & \cite{SIHosseini:09} \\ \hline
\rowcolor{lpcvdcol}
Si$_3$N$_4$ & High & No & LPCVD & 780 & 485 & N/A (passive) & \cite{SIDoolin:15} \\ \hline
\rowcolor{lpcvdcol}
Si$_3$N$_4$ & High & No & LPCVD & 780 & 26 & N/A (passive) & \cite{SISinclair:20} \\ \hline
\rowcolor{lpcvdcol}
Si$_3$N$_4$ & High & No & LPCVD & 780 & 73 & N/A (passive) & \cite{SIZhao:20} \\ \hline
\rowcolor{lpcvdcol}
Si$_3$N$_4$ & High & No & LPCVD & 925 & 55 & N/A (passive) & \cite{Buzaverov:23} \\ \hline
\rowcolor{lpcvdcol}
Si$_3$N$_4$ + AlN & High & No & LPCVD & 400--800 & 100--500 & N/A (passive) & \cite{3DSiNAlN:25} \\ \hline

\rowcolor{pecvdcol}
Si$_3$N$_4$ & Moderate/ High & Yes & PECVD & 445--561 & 550--820 & N/A (T-O, $P_\pi{=}0.78$\,mW, 0.93\,mW, 1.09\,mW, 1.2\,mW) & \cite{SIYong:22} \\ \hline
\rowcolor{pecvdcol}
Si$_3$N$_4$ & High & No & PECVD + LPCVD & 780 & 10--50 & N/A (passive) & \cite{mcnulty2025overcoming} \\ \hline
\rowcolor{pecvdcol}
Si$_3$N$_4$ & High & Yes & PECVD & $\sim$780 & N/A & N/A (1.5\,V$\cdot$cm) & \cite{SILe:26} \\ \hline
\rowcolor{pecvdcol}
Si$_3$N$_4$ & High & Yes & PECVD & 430--464, 466--500, 502--550, 555--600, 602--648 & 250--450, 220--280, 150--240, 130--160, 130--240 & N/A (passive) & \cite{SISacher:19b} \\ \hline
\rowcolor{pecvdcol}
Si$_3$N$_4$ & High & Yes & PECVD & 532, 780, 900 & 100--300 & N/A (passive) & \cite{SISubramanian:13} \\ \hline
\rowcolor{pecvdcol}
Si$_3$N$_4$ & High & Yes & PECVD & 600 & 51 & N/A (passive) & \cite{SIRomeroGarcia:13} \\ \hline
\rowcolor{pecvdcol}
Si$_3$N$_4$ & High & Yes & PECVD (anneal-free) & 737 & 38, 51 & 22, 36 & \cite{SIDong2022} \\ \hline
\rowcolor{pecvdcol}
Si$_3$N$_4$ & High & Yes & PECVD (anneal-free) & 780 & 39 & 50.7 & \cite{SIDong:22} \\ \hline

\end{tabular}
}
 \caption{\textbf{Performance comparison of visible and near-infrared (VIS–NIR) silicon nitride photonic platforms}}
\label{table1}
\end{table}

Supplementary Table~\ref{table1} presents a comprehensive summary (to the best of our current knowledge) of the propagation loss and voltage--loss product (VLP) for silicon nitride (Si$_3$N$_4$) waveguide platforms, grouped by optical confinement (Low $\rightarrow$ High). Row colour indicates the film growth method: \colorbox{lpcvdcol}{\strut\ \ } = LPCVD, \colorbox{pecvdcol}{\strut\ \ } = PECVD, and \colorbox{thiswork}{\strut\ \ } = this work.

\textit{Low confinement}: core thickness $\lesssim 90$,nm, providing ultra-low propagation loss but weak optical confinement and large bend radii. \textit{High confinement}: core thickness $\gtrsim 150$,nm, enabling compact waveguides with small bend radii at the expense of increased scattering loss.

\textbf{CMOS-BEOL compatibility} combines two independent criteria; an entry is marked \textit{Yes} only if both are satisfied. (i) \textit{Material compatibility}: the constituent materials are free of contaminants prohibited in CMOS fabrication lines. Si$_3$N$_4$ is a standard CMOS-compatible material, whereas the Pb contained in lead zirconate titanate (PZT) phase shifters poses a contamination risk. (ii) \textit{BEOL thermal budget}: all processing temperatures remain $\leq 400,^\circ$C, enabling integration on completed CMOS interconnect stacks. A platform is classified as \textit{No} if either criterion is not met. For example, LPCVD Si$_3$N$_4$ is material-compatible but typically requires a $\sim 1200,^\circ$C densification anneal that exceeds the BEOL thermal budget, while PZT-based platforms fail both criteria because of Pb contamination and the $\sim 600$--$700,^\circ$C crystallization anneal required for PZT.

Active tuning mechanisms include thermo-optic (T-O), stress-optic PZT (piezoelectric, including PZT-MEMS), and liquid-crystal (LC) phase control. LPCVD: low-pressure chemical vapor deposition; PECVD: plasma-enhanced chemical vapor deposition. Passive entries report only propagation loss, whereas active entries additionally report a VLP or an equivalent figure of merit (e.g., V$\cdot$m or $\pi$-shift power, $P_\pi$).


\section{Comparison table for active visible wavelength modulator platforms}
\label{Table2}
\setlength{\arrayrulewidth}{1.5pt}
\begin{table}[ht!]
\centering
\resizebox{\textwidth}{!}{\large
\begin{tabular}{|p{2.5cm}|p{2.6cm}|p{1.5cm}|p{2cm}|p{2.5cm}|p{2cm}|p{2cm}|p{1.5cm}|p{2.5cm}|p{2cm}|}
\hline
\textbf{Material} & \textbf{Confinement} & \textbf{$\lambda$ (nm)} & \textbf{Prop. Loss (dB/m)} & \textbf{$V_{\pi} L$ (V$\cdot$m) / $P_{\pi}$} & \textbf{VLP ($\mathrm{V}\cdot$dB)} & \textbf{BW} & \textbf{IL (dB)} & \textbf{Footprint} & \textbf{Ref.} \\
\hline
\rowcolor{sechead}
\multicolumn{10}{|l|}{\textbf{Electro-optic (Pockels effect)}} \\ \hline
\rowcolor{eocol}
TF-LiNbO$_3$ ($x$/$z$-cut) & High & 450--938 & 6--2200 & 0.0016--0.0126 & 0.096--24 & 2.7--35\,GHz & 3--8 & 3--10\,mm MZI & \cite{Renaud2023,Celik2022,CNF2023,Li:22,Desiatov:19} \\ \hline
\rowcolor{eocol}
TF-LiTaO$_3$ & High & 737 & $\sim$50 & 0.0065 & $\sim$0.33 & 20\,GHz & 5.3 & mm-scale MZI & \cite{Powell2025LT} \\ \hline
\rowcolor{eocol}
AlN on SiO$_2$ & High & 411, 766--780 & $>$1000 & $\sim$0.24--1.78 & N/A & $\sim$4\,GHz & N/A & 5--20\,mm MZI & \cite{Pernice2012,Zhu2020AlN} \\ \hline
\rowcolor{eocol}
EO polymer & High & 640 & 260 & 0.0052 & $\sim$1.35 & N/A & N/A & sub-mm MZI & \cite{Kamada:22} \\ \hline
\rowcolor{sechead}
\multicolumn{10}{|l|}{\textbf{Piezo-optomechanical (stress-optic)}} \\ \hline
\rowcolor{piezocol}
Si$_3$N$_4$ + AlN (high-conf.) & High (CMOS compatible) & 700--780 & 30--100 & 1.1--1.3 & $\sim$22--50.7 & 120\,MHz & 3.5 & $\sim$0.065\,mm$^2$ & \cite{SIDong2022,SIDong:22} \\ \hline
\rowcolor{thiswork}
\textbf{Si$_3$N$_4$ + AlN (low-conf.)} & \textbf{Low} (CMOS compatible) & \textbf{780} & \textbf{2.6--7.2} & $\sim$\textbf{2.8} & \textbf{$\sim$20.2} & \textbf{$\sim$7\,MHz} & \textbf{$\sim$4.3} & \textbf{4.84\,mm$^2$} & \textbf{This work} \\ \hline
\rowcolor{piezocol}
Si$_3$N$_4$ (hybrid electrostatic--piezo) & High (CMOS compatible) & $\sim$780 & N/A & 0.015 & $\sim$0.45--1.5 & $>$20\,MHz & N/A & 0.1\,mm$^2$ & \cite{SILe:26} \\ \hline
\rowcolor{piezocol}
Si$_3$N$_4$ + PZT (MEMS bridge) & High (non-CMOS-compatible) & 635 & $<$75 & 0.0225 & $<$1.7 & $\sim$200\,Hz & N/A & 3--5\,mm arm & \cite{Dullo2026} \\ \hline
\rowcolor{piezocol}
Si$_3$N$_4$ + PZT (coil/ring) & Low (non-CMOS-compatible) & 493--780 & 24--27 & 2.8\,V ($V_{\pi}$) & $\sim$2.34--6.71 & $\sim$0.4\,MHz & N/A &  mm-scale & \cite{SInick26_Blumenthal} \\ \hline
\rowcolor{piezocol}
Si$_3$N$_4$ + PZT & Low (non-CMOS-compatible) & 640 & N/A & 300\,nW ($P_\pi$) & N/A & $\sim$1\,MHz & N/A & 8--10\,mm & \cite{Hosseini:15} \\ \hline
\rowcolor{sechead}
\multicolumn{10}{|l|}{\textbf{MEMS}} \\ \hline
\rowcolor{memscol}
Si$_3$N$_4$ + SOI & High & 520--560 & $\sim$300 & 0.5\,nW & N/A & $\sim$30\,kHz & 2.5--3.5 & $\sim$0.7\,mm$^2$ & \cite{SIGovdeli:25} \\ \hline
\rowcolor{memscol}
Si$_3$N$_4$ & High & 795 & N/A & $<$75\,$\mu$W & N/A & 1--2\,MHz & N/A & N/A & \cite{Gyger:21} \\ \hline
\rowcolor{sechead}
\multicolumn{10}{|l|}{\textbf{Thermo-optic (resistive heater)}} \\ \hline
\rowcolor{thermocol}
Si$_3$N$_4$ & Moderate/ High & 445--561 & 550--820 & 0.78--1.3\,mW & N/A & $\sim$kHz & 2.4--6.5 & mm-scale & \cite{SIYong:22} \\ \hline
\rowcolor{thermocol}
Si$_3$N$_4$ & High & 473 & 1500 & 20--30\,mW & N/A & $\sim$50\,kHz & N/A & mm-scale & \cite{mohanty2020reconfigurable} \\ \hline
\rowcolor{thermocol}
Si$_3$N$_4$ & High & 488--530 & 260--960 & 0.85--1.8\,mW & N/A & 290\,kHz & 0.61--0.87 & $\sim$314\,$\mu$m$^2$ (microring) & \cite{liang2021robust} \\ \hline
\rowcolor{thermocol}
Si$_3$N$_4$ & High & 532 & $\sim$300 & 0.63\,mW & N/A & $\sim$kHz & N/A & mm-scale & \cite{SIWangTO:24} \\ \hline
\rowcolor{sechead}
\multicolumn{10}{|l|}{\textbf{Acousto-optic (GHz resonant)}} \\ \hline
\rowcolor{aocol}
Si$_3$N$_4$ + AlN & High & 730 & N/A & 0.0026 (@ 2.31\,GHz) & N/A & $\sim$2.31\,GHz & N/A & $\sim$2\,mm & \cite{Freedman2025} \\ \hline
\rowcolor{sechead}
\multicolumn{10}{|l|}{\textbf{Liquid crystal on SiN}} \\ \hline
\rowcolor{lccol}
Si$_3$N$_4$ + nematic LC & High & 632--780 & N/A & 0.00006--0.00014\,/ $\sim\mu$W & N/A & $<$kHz & N/A & 24--500\,$\mu$m & \cite{Notaros2022LC,Majumdar2025LC} \\ \hline
\end{tabular}
}
\caption{\textbf{Performance comparison of active visible wavelength integrated optical modulator platforms.} MZI: Mach--Zehnder interferometer. BW: modulation bandwidth. IL: insertion loss. VLP: voltage--loss product. T-O: thermo-optic. Bold row indicates this work.}
\label{table2}
\end{table}

Supplementary Table~\ref{table2} compares active visible-wavelength modulator and phase-shifter platforms, grouped by the physical tuning mechanism: electro-optic (Pockels effect) \colorbox{eocol}{\strut\ \ }, piezo-optomechanical (stress-optic) \colorbox{piezocol}{\strut\ \ }, electrostatic MEMS \colorbox{memscol}{\strut\ \ }, thermo-optic \colorbox{thermocol}{\strut\ \ }, acousto-optic \colorbox{aocol}{\strut\ \ }, and liquid-crystal \colorbox{lccol}{\strut\ \ }; the bold row \colorbox{thiswork}{\strut\ \ } is this work. The efficiency metric is mechanism-dependent: for the electro-optic and piezo-optomechanical platforms it is the voltage--length product $V_{\pi}L$, whereas the thermo-optic, MEMS, and liquid-crystal platforms are characterized by the $\pi$-shift power $P_\pi$ (or static hold power). The voltage--loss product (VLP), defined as the product of the $\pi$-shift drive voltage and the associated phase-shifter loss, captures the efficiency--loss trade-off for the field-driven (electro-optic and piezoelectric) modulators and is not applicable to the power-driven mechanisms. BW denotes the modulation bandwidth, IL the insertion loss, and confinement follows the convention of Supplementary Table~\ref{table1}. For all metrics except bandwidth, lower values are preferable.

\section{Strain-induced change in effective refractive index and estimation of photoelastic coefficients of silicon nitride }

\begin{figure*}[ht!]
\centering\includegraphics[width=\linewidth]{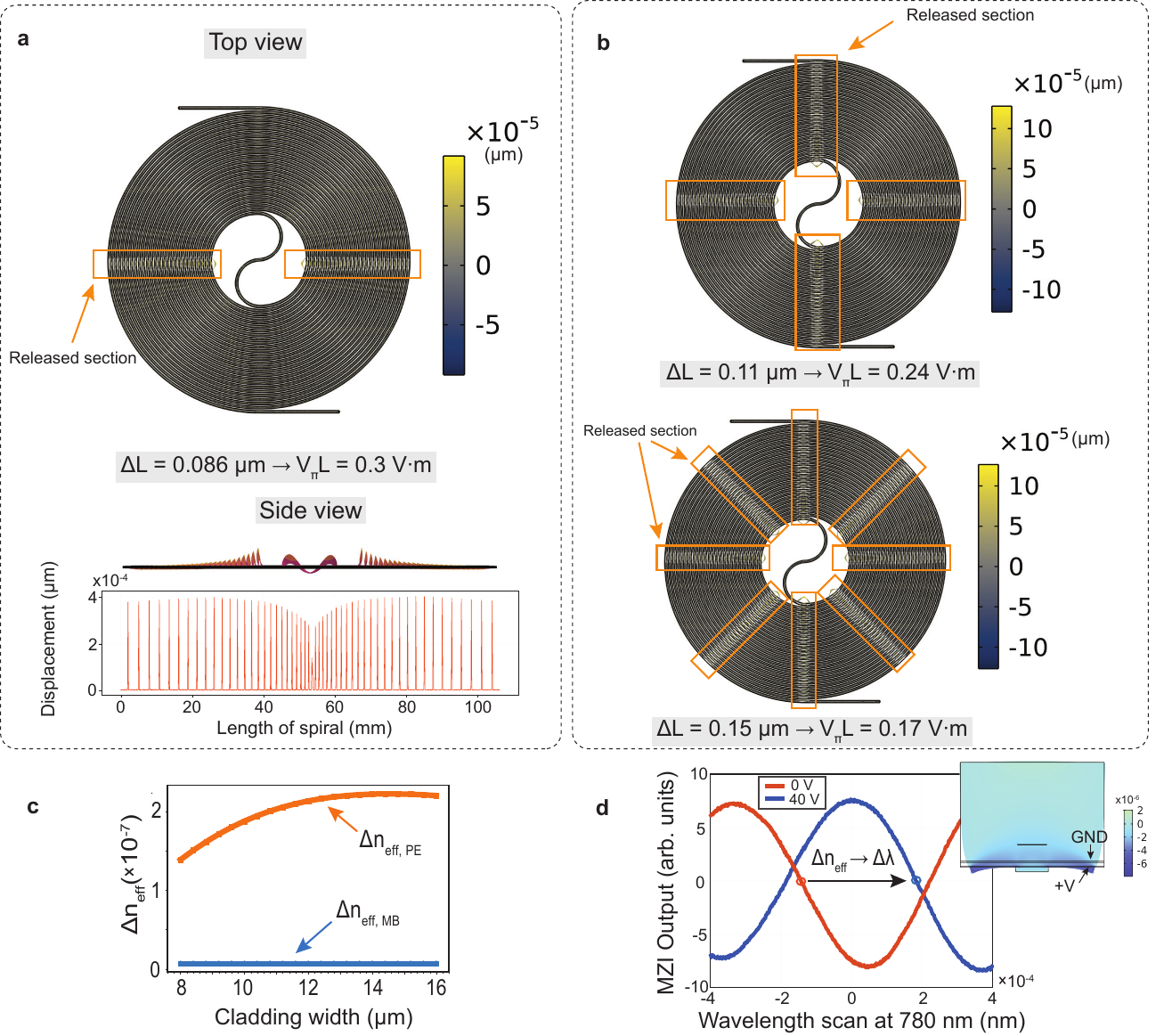}
\caption{\textbf{Strain-induced deformation and photoelastic refractive-index modulation in SPS waveguides: schematics, simulations, and wavelength shift experimental result}. \\ \textbf{(a)} Top- and side-view schematics illustrate the (exaggerated) strain-induced deformation of the released SPS section under an applied voltage. The accompanying plot presents the simulated displacement of the cladding edge along the spiral length. \textbf{(b)} A top-view schematic illustrates the exaggerated strain-induced deformation of the 4- and 8-released SPS sections under an applied voltage. The $V_\pi L$ reaches values as low as 0.17 V$\cdot$m. \textbf{(c)} FEM simulation of the strain-induced change in effective refractive index, showing that the photoelastic (PE) contribution ($\sim10^{-7}/\mathrm{V}$) exceeds the moving boundary (MB) contribution ($\sim10^{-9}/\mathrm{V}$) for different cladding widths. \textbf{(d)} Measured fringe shift when 40 V is applied to the SPS in MZI while scanning the laser wavelength at 780 nm, indicating a positive effective refractive index change under negative strain. This is possible only for the case when the dominant photoelastic contribution in these low-confinement waveguides comes from the silicon oxide cladding.
\label{Fig_ap-3VpiL}
 }
\end{figure*}

The phase shift $\Delta \theta$ experienced by an optical mode propagating through a phase shifter of length $L$ under an applied strain can be expressed as

\begin{equation} 
\Delta \theta = \frac{2\pi}{\lambda} ( \Delta n_{\mathrm{eff}} L + n_{\mathrm{eff}} \Delta L ), 
\end{equation}

where $\lambda$ denotes the free-space wavelength, $n_{\mathrm{eff}}$ is the effective refractive index, and $\Delta L$ is the strain-induced change in the optical path length \cite{SIDong2022}. In this low-confinement waveguide platform, the contribution from $\Delta L$ is negligible compared to the refractive index modulation. If, however, a few sections of the SPS are fully released, $V_{\pi} L$ could in principle be reduced by a factor of $5-10$ compared to the value measured in our current devices as shown in Supplementary Figure~\ref{Fig_ap-3VpiL}a and \ref{Fig_ap-3VpiL}b. The parameter $\Delta n_{\mathrm{eff}}$ encapsulates the total strain-induced change in effective refractive index, expressed as

\begin{equation} 
\Delta n_{\mathrm{eff}} = \Delta n_{\mathrm{eff,PE}} + \Delta n_{\mathrm{eff,MB}}. 
\end{equation}

Here, $\Delta n_{\mathrm{eff,PE}}$ quantifies the photoelastic effect, and $\Delta n_{\mathrm{eff,MB}}$ quantifies the moving boundary effect. FEM simulations for our platform reveal that the photoelastic contribution is on the order of $10^{-7}\mathrm{V^{-1}}$, whereas the moving boundary contribution is on the order of $10^{-9}\mathrm{V^{-1}}$ as shown in Fig. \ref{Fig_ap-3VpiL}c. 

The strain-induced change in the refractive index tensor component $\Delta n_{ij}$ is related to the strain tensor components $s_{kl}$ as
\begin{equation} 
\label{eq4} 
\Delta n_{ij} \propto -\sum_{k,l} p_{ijkl} s_{kl}. 
\end{equation}
Here $p_{ijkl}$ is the photoelastic tensor. An applied electric field in aluminum nitride induces strain according to 
\begin{equation} 
\label{eq5} 
s_{kl} = \sum_m d_{klm} E_m. 
\end{equation}
The relevant AlN piezoelectric coupling tensor components are $d_{33} \approx 5~\mathrm{pm/V}$, $d_{31} \approx -2~\mathrm{pm/V}$ \cite{SIStanfield19_09}, and $d_{15} \approx -2~\mathrm{pm/V}$. In these devices, the $d_{31}$ component primarily determines the piezoelectric response, as the actuator’s electric fields are nearly vertical ($E_m \approx \Delta V / t_{\mathrm{AlN}}$, with $t_{\mathrm{AlN}}$ being the AlN layer thickness).

\begin{figure*}[ht!]
\centering\includegraphics[width=\linewidth]{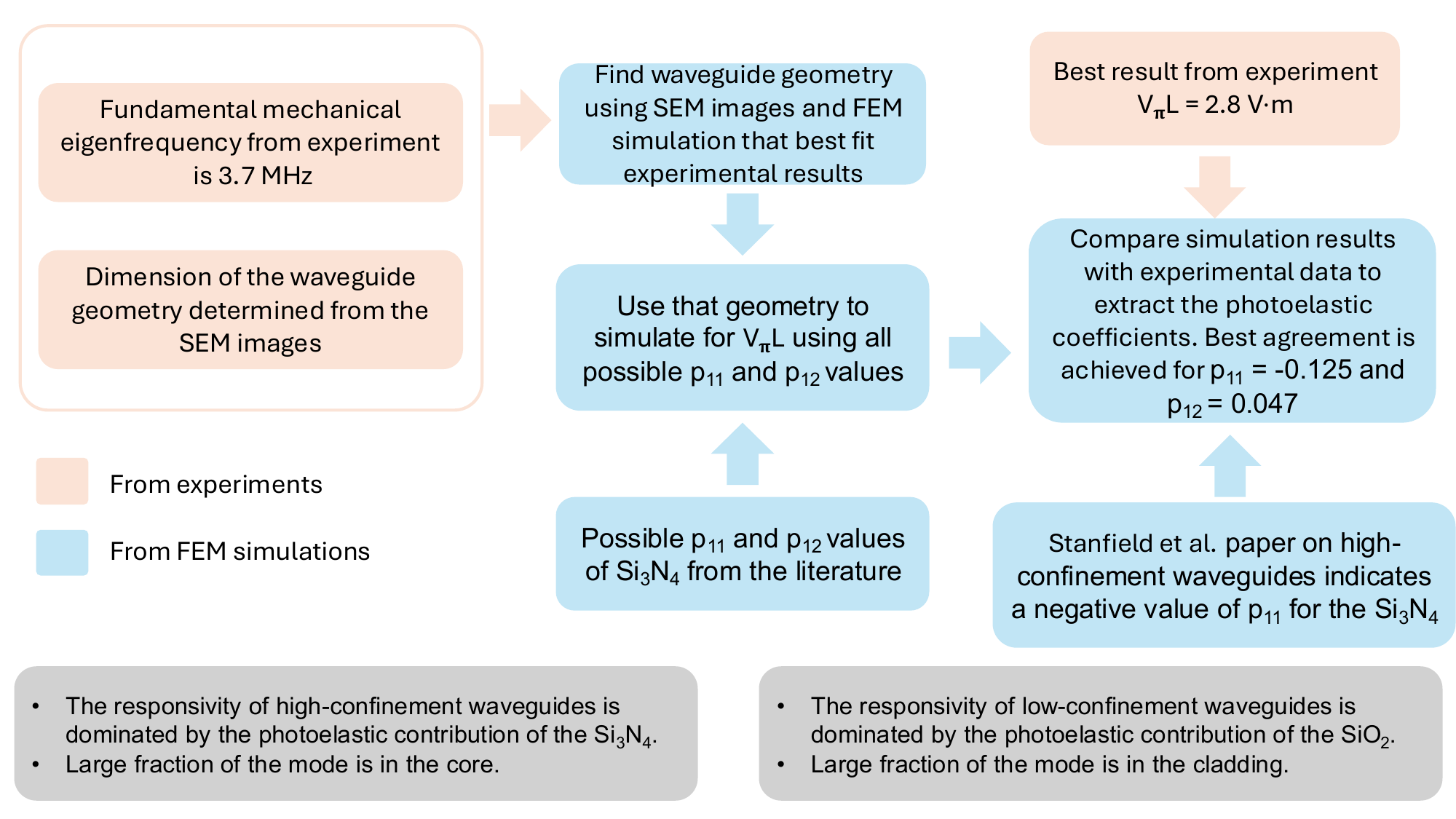}
\caption{\textbf{Simulation workflow for the estimation of silicon nitride photoelastic coefficients.} \\ The simulation incorporates actual waveguide dimensions to identify the best-fit parameters $p_{11} = -0.125$ and $p_{12} = 0.047$.
\label{work_flow}
}
\end{figure*}

\subsection{Rough estimate of silicon nitride photoelastic coefficients $p_{11} = -0.125$ and $p_{12} = 0.047$}

\label{SubappendixVpiL1}

\begin{figure*}[ht!]
\centering\includegraphics[width=\linewidth]{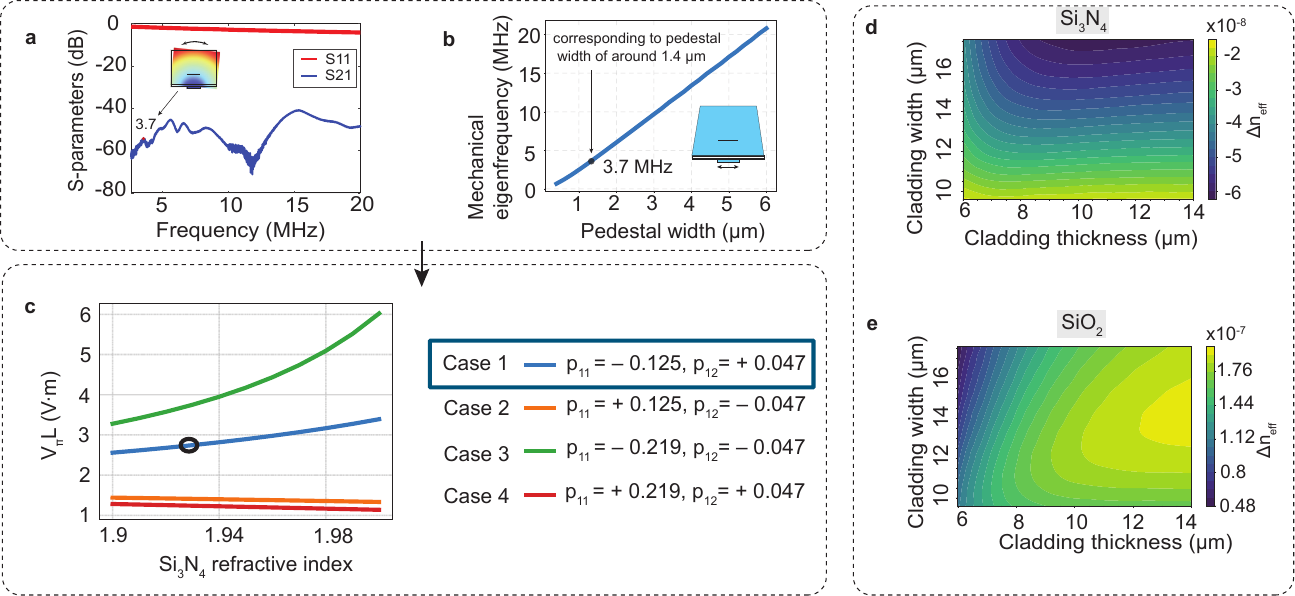}
\caption{\textbf{Photoelastic coefficient determination through mechanical, optical characterization and FEM simulation.}\\
\textbf{(a)} Measured S${11}$ and S${21}$ responses of the SPS, showing a fundamental mechanical resonance at approximately 3.7~MHz and for the same device $V_{\pi}L\sim2.8$.  \textbf{(b)} FEM-derived relationship between waveguide pedestal width and mechanical eigenfrequency; a resonance at 3.7~MHz corresponds to a pedestal width of $\approx1.4~\mu$m. \textbf{(c)} FEM-simulated $V_{\pi} L$ values for geometry with pedestal width of $\approx1.4~\mu$m, evaluated for four combinations of silicon nitride photoelastic coefficients and refractive indices. The parameter set $p_{11} = -0.125$ and $p_{12} = 0.047$ provides the best agreement with the experimental data for $\mathrm{Si}_{3}\mathrm{N}_{4}$ refractive index of 1.93. \textbf{(d)} Simulated change in effective refractive index per applied voltage using silicon nitride photoelastic coefficients ($p_{11} = -0.125$, $p_{12} = 0.047$) with silicon dioxide photoelastic coefficients set to zero, resulting a negative $\Delta n_{\mathrm{eff}}/V$. \textbf{(e)} Corresponding simulation using only silicon dioxide photoelastic coefficients ($p_{11} = 0.121$, $p_{12} = 0.27$), showing a positive and an order of magnitute larger $\Delta n_{\mathrm{eff}}/V$, indicating a dominant PE effect in SiO$_2$ and an opposing influence from Si$_3$N$_4$.
\label{Fig_ap-2VpiL}
 }
\end{figure*}

The photoelastic coefficients used in our initial simulation while designing devices and in results shown in Supplementary Figure~\ref{Fig_ap-3VpiL}c are as follows: for silicon nitride, $p_{11} = 0.1$ (assumed), $p_{12} = 0.047$ (assumed) \cite{SIGyger2020a, SITian:24}, and $p_{44} = 0.5(p_{11}-p_{12}) = 0.0265$; for silicon dioxide, $p_{11} = 0.121$, $p_{12} = 0.27$, and $p_{44} = 0.5(p_{11}-p_{12}) = -0.0745$ \cite{SITian:24}. In the FEM simulations, we use a refractive index of 1.92 for silicon nitride and 1.45 for silicon dioxide.

The reported values for silicon nitride photoelastic coefficients are $|p_{12}| = 0.047 \pm 0.004$ and $|p_{44}| = 0.086$ \cite{SIGyger2020a, SITian:24}. Based on these values, there are four possible sign combinations for $p_{11}$, corresponding to the two magnitudes $|0.125|$ and $|0.219|$. 

We estimate the waveguide cross-sectional dimensions from the SEM images in Fig.~\ref{Fig2}a and Fig.~\ref{Fig2}b in main text. We extract the pedestal width of $1.4~\mu\mathrm{m}$ using mechanical resonance measurements of the same device that we use to report $V_{\pi} L$ in the main text (see Supplementary Figure~\ref{Fig_ap-2VpiL}a) and the FEM simulation (see \ref{Fig_ap-2VpiL}b). Using these geometric parameters, we run FEM simulations to compute $V_{\pi} L$ for the four cases, shown in Supplementary Figure~\ref{Fig_ap-2VpiL}c. Among them, only the combination $p_{11} = -0.125$ and $p_{12} = 0.047$ of silicon nitride yields values that closely match our experimental measured value of $V_{\pi}L=2.8~\mathrm{V}\cdot \mathrm{m}$. We emphasize that these are best-estimate values inferred from our measurements, intended to guide future device designs rather than serve as definitive material parameters. The workflow for this estimation procedure is summarized in Supplementary Figure~\ref{work_flow}.

%

When we use $p_{11} = -0.125$ and $p_{12} = 0.047$ for silicon nitride and set the photoelastic coefficients of silicon dioxide ($p_{11}$ and $p_{12}$) to zero, the simulated change in the effective refractive index per applied voltage, as shown in Supplementary Figure~\ref{Fig_ap-2VpiL}d, is negative with respect to changes in the cladding width and thickness. Conversely, when we set the photoelastic coefficients of silicon nitride to zero and use $p_{11} = 0.121$ and $p_{12} = 0.27$ for silicon dioxide, the change in the effective refractive index per applied voltage becomes positive and an order of magnitude larger, as shown in Supplementary Figure~\ref{Fig_ap-2VpiL}e. These results indicate that silicon dioxide exerts a dominant influence on the voltage-induced change in refractive index because a large fraction of the optical mode resides in the cladding, whereas the silicon nitride core contributes an opposite, smaller effect. This observation agrees with the experimentally measured wavelength shift of the MZI when negative strain is applied to the SPS, as shown in Supplementary Figure~\ref{Fig_ap-3VpiL}d.

\begin{figure*}[ht!]
\centering\includegraphics[width=0.8\linewidth]{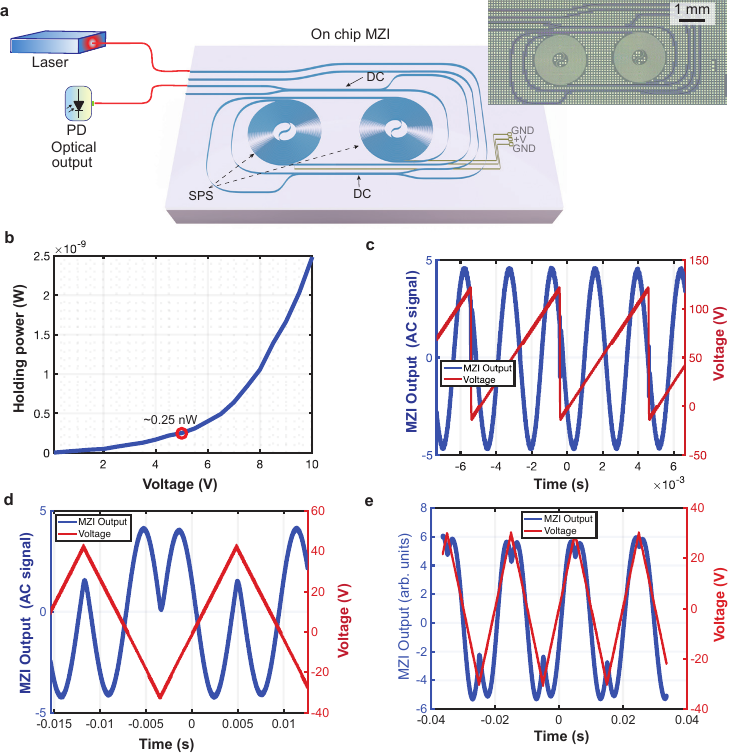}
\caption{\textbf{On-chip MZI and SPS characterization.}\\ \textbf{(a)} Schematic of the on-chip Mach–Zehnder interferometer operating in a differential configuration, with inset showing a micrograph of the fabricated device. \textbf{(b)} Measured holding power as a function of applied voltage for the SPS, showing $0.25~\mathrm{nW}$ holding power at a $5~\mathrm{V}$ actuator bias. \textbf{(c)} Optical response of the same SPS corresponding to the device reported in Fig.~\ref{Fig2} in main text.
\textbf{d, e} Optical responses of two additional MZI devices incorporating $\sim9.8~\mathrm{cm}$-long Archimedean SPSs in a push–pull configuration.
\label{Fig_ap-1VpiL}
 }
\end{figure*}

\subsection{Voltage-length product ($V_{\pi}L$) of other devices }
\label{SubappendixVpiL}

To characterize the devices, we use a function generator as voltage source in combination with a Thorlabs BPA100 benchtop high-voltage piezo amplifier, which provides a noise bandwidth of 20 Hz to 100 kHz. Supplementary Figure~\ref{Fig_ap-1VpiL}a shows the on-chip Mach–Zehnder interferometer operating in a differential configuration, with the inset displaying a micrograph of the on-chip MZI. We apply a $200~\mathrm{Hz}$ ramp modulation signal, sweeping the voltage from approximately $-20~\mathrm{V}$ to $40~\mathrm{V}$ and up to $120~\mathrm{V}$, depending on the test condition. Supplementary Figure~\ref{Fig_ap-1VpiL}b presents the holding power as a function of applied voltage ($0$–$10~\mathrm{V}$) for the SPS reported in the main text, showing a holding power of $0.25~\mathrm{nW}$ at an actuator bias of $5~\mathrm{V}$. Supplementary Figure~\ref{Fig_ap-1VpiL}c shows the same SPS response reported in the main text (Fig.~\ref{Fig2}). Supplementary Figure~\ref{Fig_ap-1VpiL}d and e display the optical responses of two additional MZI devices incorporating $\sim9.8~\mathrm{cm}$-long Archimedean SPSs in a push–pull configuration, with measured $V_{\pi}$ values of approximately $40~\mathrm{V}$.

\section{Heralded Bell state and 4-mode GHZ state generation using lossy LPNP circuit}
\label{Appendix3}
\begin{figure*}[ht!]
\centering
\includegraphics[width=\linewidth]{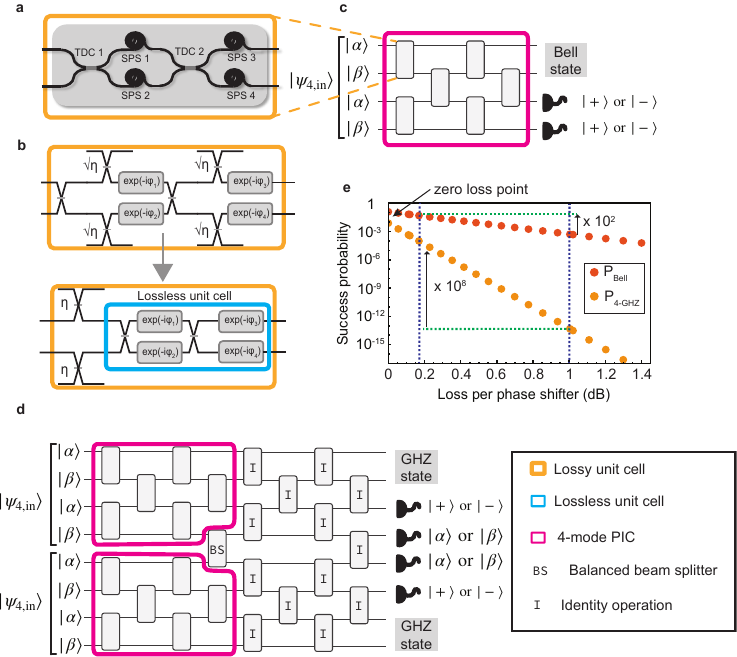}
\caption{\textbf{Impact of phase-shifter loss on heralded Bell and GHZ state generation in LPNP photonic circuits.} \\ 
\textbf{(a)} Lossy unit cell of an LPNP circuit consists of two tunable directional couplers and four spiral phase shifters. \textbf{(b)} We simulate the phase shifter losses in a lossy unit cell by employing beam splitters with transmissivity $\sqrt{\eta}$ that interact with a vacuum mode representing the environment. \textbf{c, d} show 4-mode and 8-mode PICs initialized with the input state $|\Psi_{4,\mathrm{in}}\rangle$ to herald the generation of Bell and 4-mode GHZ states. \textbf{(e)} Probability of heralding Bell and GHZ states for these protocols in the presence of losses.
\label{Fig_ap1}
}
\end{figure*}

In this calculation, we assume that the dominant loss in the unit cell (Supplementary Figure~\ref{Fig_ap1}a) of the LPNP circuit is the propagation loss of the phase shifters, while all other components are considered lossless. As described in the discussion section of the paper, we consider two loss cases of the phase shifter $\gamma = 1~\mathrm{dB}$, and $\gamma=0.2~\mathrm{dB}$. To simplify the analysis, we model the loss in each phase shifter as a beam splitter with transmissivity $\sqrt{\eta}$ on each arm. Since these beam splitters have equal losses on both arms, they commute with other beam splitters and single-mode phase shifters within the unit cell. Therefore, a lossy unit cell with beam splitters next to phase shifters is equivalent to a lossy unit cell with beam splitters concentrated at the input, as shown in Supplementary Figure~\ref{Fig_ap1}b. To convert phase shifter loss $\gamma$ into linear transmissivity, we employ the following equation:

\begin{equation}
\eta = (10^{-\gamma/10})^2.
\end{equation}
For $\gamma=0.2~\mathrm{dB}$, this yields $\eta_{\mathrm{1}} = 0.91$, while for $\gamma=1~\mathrm{dB}$, $\eta_{\mathrm{2}} = 0.63$.

We adopt the GHZ state heralding protocol outlined in~\cite{SIBhatti:25} to assess the success probability for on-demand generation of Bell and 4-mode GHZ states via a lossy PIC. For heralded Bell state generation, two pairs of indistinguishable single photons occupying orthogonal internal degrees of freedom ($D \in \{\alpha, \beta\}$) are used and passed through a 4-mode PIC, as illustrated in Supplementary Figure~\ref{Fig_ap1}c. The input state to the 4-mode PIC is given by
\begin{equation}
\begin{aligned}
|\Psi_{4,\mathrm{in}}\rangle = \prod_{k=1}^{4} a_{D,k}^{\dagger} |0\rangle 
= |1\rangle_{\alpha,1} \otimes |1\rangle_{\beta,2} \otimes |1\rangle_{\alpha,3} \otimes |1\rangle_{\beta,4} 
\equiv |\alpha\rangle_{1}|\beta\rangle_{2}|\alpha\rangle_{3}|\beta\rangle_{4}
\end{aligned}
\end{equation}
where $a_{D,k}^{\dagger}$ ($a_{D,k}$) denotes the creation (annihilation) operator for a photon in the $k$th spatial input mode with internal degree of freedom $D \in \{\alpha, \beta\}$, and $|n\rangle_{D,k}$ denotes the corresponding Fock state.

The unitary transformation $U$ maps the 4-mode input state to a 4-mode output state using a 4-port PIC, with matrix elements $U_{kl}$ ($k,l = 1,2,3,4$), mapping input creation operators $a_{D,k}^{\dagger}$ to output creation operators $b_{D,l}^{\dagger}$ as follows:

\begin{equation}
a_{D,k}^{\dagger} \rightarrow \sum_{l=1}^{4} U_{kl} b_{D,l}^{\dagger},
\label{eq5_ap1}
\end{equation}

\begin{equation}
U_{kl} = \frac{1}{\sqrt{4}} \omega^{(k-1)(l-1)}.
\label{eq6_ap1}
\end{equation}

where $\omega = \exp(i2\pi/4)$. To account for loss within the unitary transformation, we decompose the unitary matrix into a Clements mesh~\cite{SIClements:16} and introduce equal loss to each arm of the unit cell, as illustrated in Supplementary Figure~\ref{Fig_ap1}b.

Heralding of Bell states in the first and second modes is achieved by performing temporal mode-resolving photon-number-resolving detection (PNRD)~\cite{SIBhatti:25} on the third and fourth output modes. The specific click patterns associated with these heralded Bell states are:
\begin{equation}
|+\rangle_3|+\rangle_4 \text{ and } |-\rangle_3|-\rangle_4 \rightarrow \frac{1}{\sqrt{2}} (|\alpha\rangle_1|\alpha\rangle_2 - |\beta\rangle_1|\beta\rangle_2)
\label{eq7_ap1}
\end{equation}
\begin{equation}
|+\rangle_3|-\rangle_4 \text{ and } |-\rangle_3|+\rangle_4 \rightarrow \frac{1}{\sqrt{2}} (|\alpha\rangle_1|\alpha\rangle_2 + |\beta\rangle_1|\beta\rangle_2)
\label{eq8_ap1}
\end{equation}
where $|+\rangle = (|\alpha\rangle + |\beta\rangle)/\sqrt{2}$ and $|-\rangle = (|\alpha\rangle - |\beta\rangle)/\sqrt{2}$ are the rotated basis states used for mode-resolving measurements. In the lossless limit, the probability of success is 0.125, as shown in Supplementary Figure~\ref{Fig_ap1}e. With the detection of the click patterns shown in Supplementary Equation~\ref{eq7_ap1} and \ref{eq8_ap1}, a Bell state in the corresponding form is successfully heralded with fidelity of 1. We evaluated the probability of success for this protocol in the presence of losses, utilizing the models depicted in Supplementary Figure~\ref{Fig_ap1}b and \ref{Fig_ap1}c, in which we monitor the same specified click pattern. The probability of success of the protocol in the presence of losses is shown in Supplementary Figure~\ref{Fig_ap1}e. This protocol can be scaled up by stacking larger-mode PICs and performing the corresponding measurements at the appropriate output ports.

For example, the heralded 4-mode GHZ state preparation protocol employs 8-mode PICs, initialized with two $|\Psi_{4,\mathrm{in}}\rangle$ states. The two 4-mode PICs in the 8-mode PIC are interconnected via a balanced beam splitter to interfere the fourth and fifth modes, as shown in Supplementary Figure~\ref{Fig_ap1}d. By applying Clements' decomposition to the unitary transformation, we determine the contribution of each lossy unit cell within the 8-mode PIC, half of which are identity operations. Heralding a 4-mode GHZ state in the first, second, seventh, and eighth output ports involves specific measurements on the third, fourth, fifth, and sixth modes, performed in the original and rotated bases, respectively. In the ideal case without losses, this state can be heralded with a probability of $1/128$, with a deterministic click pattern indicating successful preparation (as detailed in~\cite{SIBhatti:25}), while Supplementary Figure~\ref{Fig_ap1}e illustrates the heralded probability in the presence of losses.

\section{Propagation loss measurements}
\label{Propagation_loss}

\subsection{Top-down imaging measurement}
\label{Top_down}

Fig.~\ref{Fig4}a in main text presents the experimental setup for measuring waveguide loss using the top-down imaging technique. We capture images with an Allied Vision 1800 U-120c camera (1,280 × 960 pixels), equipped with a Navitar $12\times$ UltraZoom Video Microscopy System and a Mitutoyo $5\times$ microscope objective (MO). Because the low-confinement waveguides exhibit very low loss and therefore minimal light scattering, we adjust the imaging parameters during acquisition to improve the signal-to-noise ratio. For the unetched waveguide, where we image a section of the long spiral $1.8~\mathrm{m}$, we use the following postprocessing parameters: gamma correction = 0.5, gain = $35~\mathrm{dB}$, and exposure time = $8.03~\mathrm{ms}$. For the etched waveguide with actuator, we set the parameters to gamma correction = 0.5, gain = $38.3~\mathrm{dB}$, and exposure time = $7.3~\mathrm{ms}$.

We extract the propagation loss by performing a linear fit of the logarithm of the scattered intensity as a function of propagation distance. For the actuator-free waveguide, this yields a propagation loss of $2.57 \pm 0.38~\mathrm{dB/m}$ with a $95\%$ confidence interval of $[1.79,~3.35]~\mathrm{dB/m}$ and an RMSE of $0.35~\mathrm{dB}$. For the actuator-integrated waveguide, we obtain a propagation loss of $7.23 \pm 0.76~\mathrm{dB/m}$ with a $95\%$ confidence interval of $[5.72,~8.75]~\mathrm{dB/m}$ and an RMSE of $0.18~\mathrm{dB}$. The relatively large residuals reflect the inherently low signal-to-noise ratio of the scattered intensity in ultra-low-loss waveguides buried beneath a thick oxide cladding: at longer propagation distances, where the guided power approaches the camera noise floor, the logarithmic conversion amplifies residual noise and increases scatter in the fitted residuals. These top-down values are independently cross-verified by OFDR measurements, which yield approximately $3~\mathrm{dB/m}$ for the actuator-free waveguide and
approximately $8~\mathrm{dB/m}$ for the actuator-integrated waveguide, in close agreement with the top-down estimates.

\begin{figure*}[ht!]
\centering
\includegraphics[width=\linewidth]{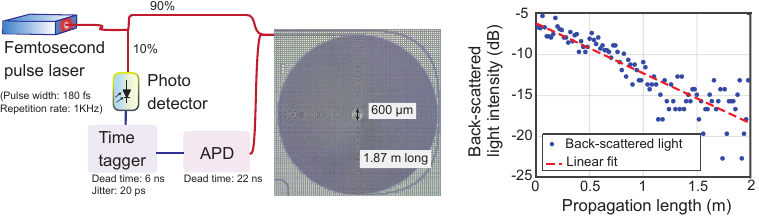}
\caption{\textbf{Waveguide loss measurements using optical time-domain reflectometry (OTDR).} \\ 
OTDR for loss characterization, utilizing a femtosecond laser, a time-tagger with a fast detector for pulse triggering, and an avalanche photodetector (APD). Linear fit indicates a loss of $6.1~\mathrm{dB/m}$.}
\label{Fig_OTDR}
\end{figure*}

\subsection{Optical time-domain reflectometry (OTDR)}
\label{OTDR}
The experiment employs femtosecond laser-based optical time-domain 
reflectometry \cite{SINatCom:22} to characterize propagation loss in a 
$1.8~\mathrm{m}$ silicon nitride spiral waveguide. A Ti:sapphire femtosecond 
laser operating at $800~\mathrm{nm}$ generates $180~\mathrm{fs}$ pulses at a 
$1~\mathrm{kHz}$ repetition rate. The laser light is injected into the 
waveguide by first coupling the free-space beam into a single-mode fiber, 
which is then coupled to the waveguide using a fiber array. Backscattered 
photons from Rayleigh scattering and localized defects are detected using an 
avalanche photodiode (APD) with a $22~\mathrm{ns}$ dead time. Timing is 
synchronized by a fast photodetector that triggers a time tagger on each laser 
pulse. Photon counts are converted to optical intensity (on a $\mathrm{dB}$ 
scale) and plotted against round-trip time, which is linearly proportional to 
the distance along the waveguide. A linear fit to the $0$--$0.9~\mathrm{m}$ 
region quantifies the propagation loss coefficient as approximately 
$6.1~\mathrm{dB/m}$, as shown in Supplementary Figure~\ref{Fig_OTDR}b. This is because 
bending-induced loss increases significantly near the spiral center, reached 
after approximately $0.9~\mathrm{m}$ of one-way propagation.

The practical spatial resolution of the OTDR measurement is 
not set by the femtosecond pulse width but by the timing constraints of the 
detection electronics: the APD dead time ($22~\mathrm{ns}$), the time-tagger 
dead time ($6~\mathrm{ns}$), and the time-tagger jitter ($20~\mathrm{ps}$), 
yielding a minimum resolvable round-trip distance of approximately 
$2.25~\mathrm{mm}$ ($\Delta L = c \cdot \tau_{\text{dead}} / 2n_g$). This $2.25~\mathrm{mm}$ resolution is sufficient for 
characterizing the $1.8~\mathrm{m}$ spiral waveguide but prevents reliable 
loss extraction from actuator-integrated waveguides shorter than approximately 
$10~\mathrm{cm}$, as the device length spans only a small number of resolvable 
spatial bins and reflections from the input and output facets dominate the 
usable measurement window.

For the spiral waveguide, the OTDR fit is restricted to the 
$0$--$0.9~\mathrm{m}$ region of the one-way propagation distance. Since the 
OTDR measures the round-trip signal, this corresponds to a two-way light path 
of $0$--$1.8~\mathrm{m}$, which is why the time-domain trace extends to 
approximately $2~\mathrm{m}$ in the plot. Close to the center of the spiral at 
around $0.9~\mathrm{m}$ of one-way propagation, the minimum bending radius 
(${\sim}150~\mathrm{\mu m}$) causes significant radiation loss that attenuates 
the backscattered signal below the reliable detection threshold, preventing 
further loss extraction beyond this point. The extracted value of 
$6.1~\mathrm{dB/m}$ is therefore higher than the $2.57~\mathrm{dB/m}$ 
obtained from top-down imaging, not due to a true discrepancy in the intrinsic 
waveguide loss, but because the OTDR fit is performed over a region that 
includes sections of progressively decreasing bend radius, where increasing 
radiation loss elevates the fitted slope. This interpretation is confirmed by 
OFDR measurements, which spatially resolve the large-bend-radius and 
small-bend-radius regions independently and cross-verify both the top-down and 
OTDR results, as discussed in the following subsection.

\subsection{Optical frequency-domain reflectometry (OFDR)}
\label{OFDR}

\begin{figure*}[ht!]
\centering
\includegraphics[width=\linewidth]{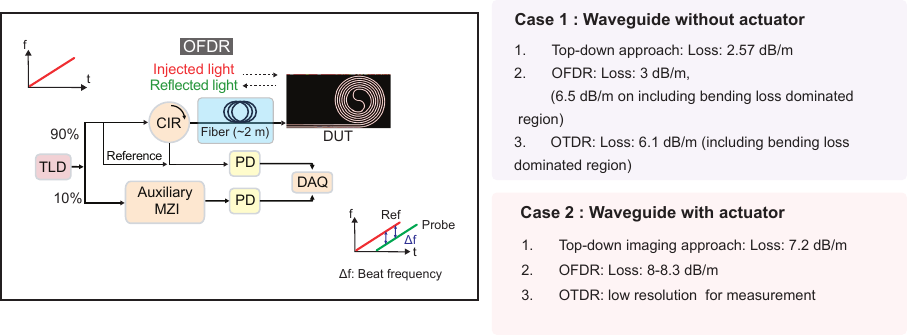}
\caption{\textbf{Waveguide loss measurement using optical frequency-domain reflectometry (OFDR).}  \\ 
A tunable laser diode (TLD) sweeps 773.2–781~nm and launches light through a circulator (CIR) into the device under test (DUT). The system interferes the backscattered signal with a reference and detects it using a photodetector (PD). An auxiliary unbalanced Mach--Zehnder interferometer calibrates the frequency sweep for Fourier analysis.}
\label{Fig_OFDR_1}
\end{figure*}

Our OFDR system maps reflections along the waveguide by sweeping a tunable laser from 773.2 nm to 781 nm (3.75 THz bandwidth) at a DAQ sampling rate of $F_s = 1$ MHz. The total optical bandwidth sets the spatial resolution $\delta L = c/(2n_g\Delta\nu) \approx 27~\mathrm{\mu m}$, while the Nyquist-limited maximum range is $L_\mathrm{Nyquist} = F_s c/(4n_g\beta) \approx 30~\mathrm{m}$, well beyond our device lengths. An auxiliary Mach–Zehnder interferometer (FSR = 50 MHz) corrects laser sweep nonlinearity by resampling the transmission signal onto a uniform optical frequency grid before Fourier transformation, ensuring each FFT bin corresponds to a unique beat frequency and therefore a unique physical distance via $L = f_\mathrm{beat} \cdot c / (2n_g\beta)$.

\begin{figure*}[ht!]
\centering
\includegraphics[width=\linewidth]{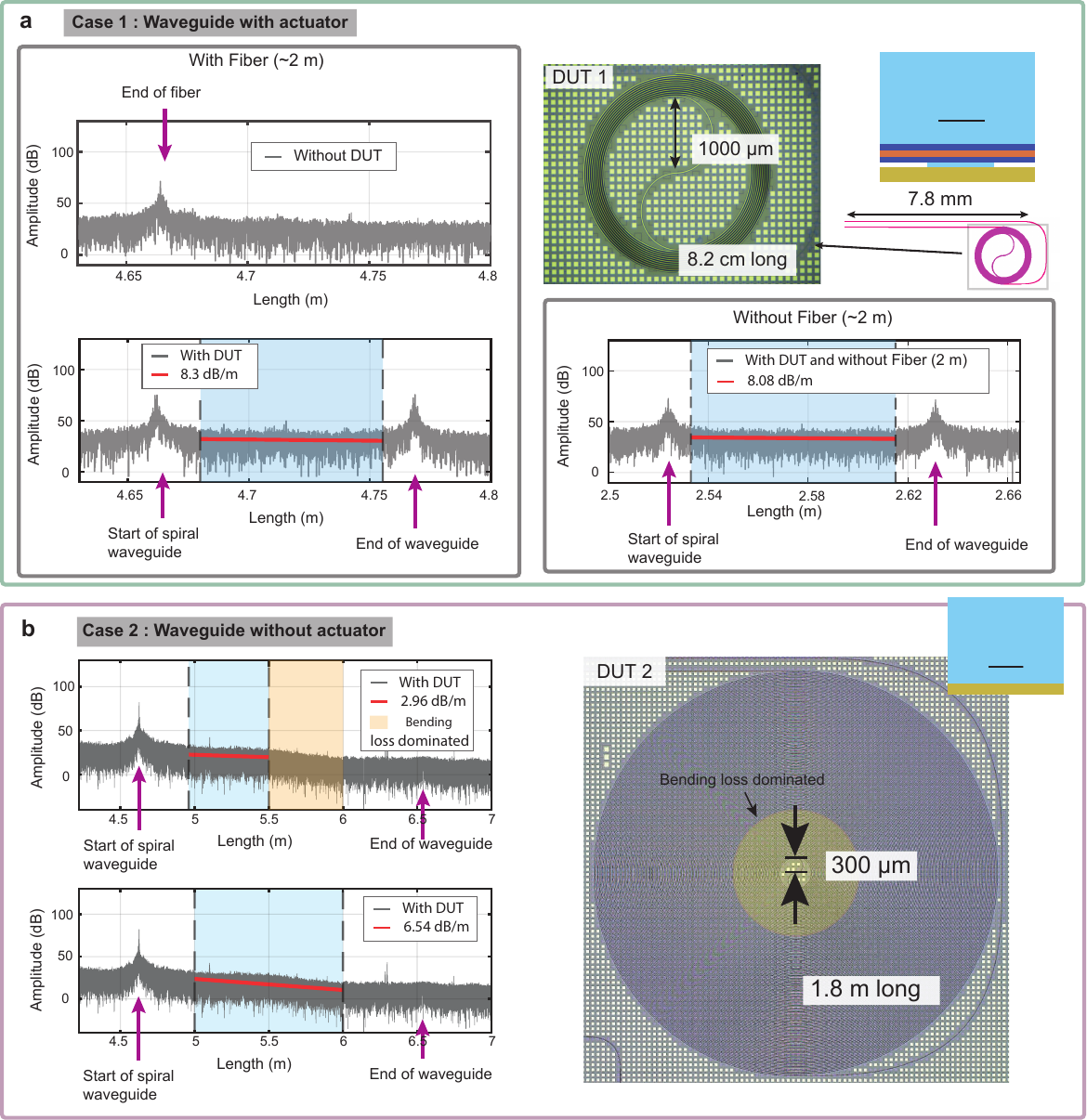}
\caption{\textbf{Waveguide loss measurements using OFDR.} (a) Loss characterization of an 8.2~cm SiN waveguide with actuator (right), with a schematic of the waveguide cross-section. The first two traces include an additional 2~m fiber in the DUT arm, showing measurements without and with the DUT, yielding $\sim 8.3$~dB/m; the third trace excludes the extra fiber and gives $\sim 8.1$~dB/m. (b) OFDR measurement of a 1.8~m waveguide without actuator. The first trace corresponds to a large bend-radius region with $\sim 3$~dB/m loss, while the second includes bend-dominated regions with $\sim 6.5$~dB/m loss. }
\label{Fig_OFDR_2}
\end{figure*}

The OFDR results are as follows. For the actuator-free waveguide, the larger bend radius sections of the 1.8 m long spiral yield a propagation loss of approximately 3 dB/m, consistent with the 2.57 dB/m from top-down imaging. When the bending-loss-dominated regions close to the spiral centre are included in the fit, the effective loss rises to approximately 6.5 dB/m, in close agreement with the 6.1 dB/m from OTDR, confirming that the OTDR value reflects spatial averaging over the high-loss bending region rather than a higher intrinsic waveguide loss. For the actuator-integrated waveguide, OFDR yields a propagation loss of approximately 8 dB/m, consistent with the 7.2 dB/m from top-down imaging. This indicates that the actuator contributes approximately 4–6 dB/m of excess loss beyond the bare waveguide, attributed to increased scattering at the actuator–waveguide interface and cladding-air interface, and a small contribution from optical absorption by the metal layer. Taken together, top-down imaging, OTDR, and OFDR consistently yield propagation losses of 2.6–3 dB/m for the larger bend radius sections of the 1.8 m long spiral and 7–8.3 dB/m for approximately 8 cm long actuator-integrated waveguides.


\section{Loss-responsivity trade-off}
\label{Loss-reponsivity}
During the initial design stage, we assumed a strain-optic coefficient of $p_{11} = 0.1$ for silicon nitride and calculated the $V_{\pi} L$ and losses for various device configurations. Supplementary Figure~\ref{Fig_Loss-reponsivity} presents these preliminary design results. For example, in Supplementary Figure~\ref{Fig_Loss-reponsivity}c, we observe that increasing the cladding width while maintaining a fixed pedestal width of $4~\mu\mathrm{m}$ decreases the $V_{\pi} L$ value (i.e., increases the responsivity) because the strain in the waveguide increases accordingly. Although positioning the waveguide core closer to the actuator improves the responsivity (reduces $V_{\pi} L$) through enhanced strain modulation, as demonstrated in Supplementary Figure~\ref{Fig_Loss-reponsivity}e, this configuration also increases the optical absorption loss per unit length ($\alpha_{\mathrm{abs}}$) from the actuator's top metal electrode. To quantify this trade-off, we employed the same FEM simulation to calculate the product of $\alpha_{\mathrm{abs}}/L$ (induced by the $100~\mathrm{nm}$-thick aluminum layer) and the corresponding $V_{\pi} L$ value for each waveguide geometry. As depicted in Supplementary Figure~\ref{Fig_Loss-reponsivity}e, the product $V_{\pi}L \cdot \alpha_{\mathrm{abs}}/L$ serves as a useful indicator to approximate an optimal core height ($h$) of around $2.5~\mathrm{\mu m}$, which minimizes the contributions of both $V_{\pi} L$ and $\alpha_{\mathrm{abs}}/L$. Although the product $V_{\pi}\alpha_{\mathrm{abs}}$ provides a starting point for optimizing the core-actuator separation, it is essential to acknowledge other factors, such as waveguide geometry, which can have a more significant influence on both $V_{\pi} L$ and $\alpha_{\mathrm{abs}}/L$.

\begin{figure*}[ht!]
\centering
\includegraphics[width=\linewidth]{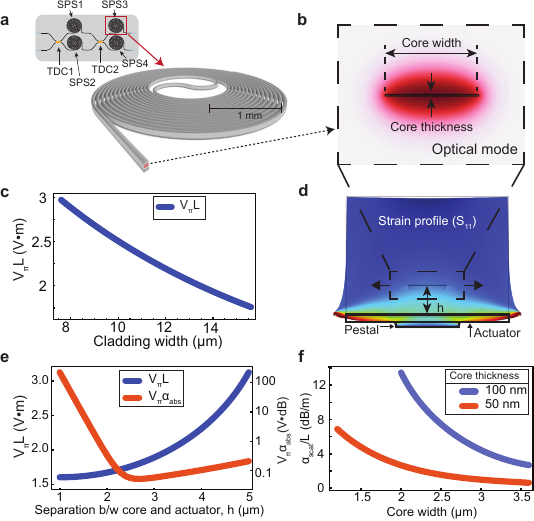}
\caption{\textbf{Design and performance optimization of the piezo-optomechanical phase shifter.} \\
\textbf{(a)} The basic building block of an MZM, accompanied by a magnified image of the SPS, which has a $1~\mathrm{mm}$ minimum bend radius.
\textbf{(b)} Cross-sectional view of the fundamental transverse electric ($\mathrm{TE}_\mathrm{0}$) mode in a low-confinement waveguide, simulated using FEM, with dimensions of $50~\mathrm{nm}$ thickness and $3.6~\mathrm{\mu m}$ width.
\textbf{(c)} $V_{\pi} L$ value as a function of cladding width, with a fixed $4~\mathrm{\mu m}$ pedestal width.
\textbf{(d)} Schematic of a phase shifter's cross-section, showing the induced $s_{11}$ component of the strain tensor and waveguide deformation (exaggerated for visualization purposes).
\textbf{(e)} $V_{\pi} L$ value (blue) and $V_{\pi}\alpha_{\mathrm{abs}}$ value (orange) as functions of waveguide core height ($h$) relative to the actuator base.
\textbf{(f)} Scattering loss per unit length for waveguides with core thicknesses of $50~\mathrm{nm}$ and $100~\mathrm{nm}$ at an optical wavelength of $780~\mathrm{nm}$. FEM simulation results are shown in blue and orange, while calculations from reference~\cite{SIChen2019} are shown in green. The calculations assume a roughness variance $\sigma^2 = 5~\mathrm{nm}^2$ and correlation length $L_c = 50~\mathrm{nm}$.
}
\label{Fig_Loss-reponsivity}
\end{figure*}

\section{Scattering loss model}
\label{Scattering_loss}

Light scattering in optical waveguides primarily originates from surface roughness present at both the core–cladding and cladding–air interfaces; these imperfections are inherent to the fabrication process and disrupt the otherwise smooth dielectric boundaries, introducing localized refractive index fluctuations that scatter guided modes. 

\subsection{3D loss model}

In order to rigorously quantify optical loss due to surface roughness, as it varies with core dimensions, we utilize the volume current method (VCM) within a FEM simulation framework. The VCM enables modeling of radiative losses stemming from dielectric perturbations by treating these as equivalent current densities $\vec{J}(\vec{r})$ \cite{SIBauters:13,SIBarwicz:05,SIBauters:11,SIKuznetsov:83, SIKhalatpour2025}.  The magnitude and direction of the equivalent current density $\vec{J}(\vec{r})$ are proportional to the local electromagnetic field $\vec{E}(x, y, z)$ of the guided mode at the rough interface \cite{SIBarwicz:05}. Specifically,

\begin{equation}
    \vec{J}(\vec{r}) = -i\omega\epsilon_0(n_{\mathrm{core}}^2 - n_{\mathrm{clad}}^2)\vec{E}(x,y,z)
\end{equation}
where $\omega$ is the angular frequency of incident light, $\epsilon_0$ is the vacuum permittivity, and $n_{\mathrm{core}}$, $n_{\mathrm{clad}}$ represent the refractive indices of the waveguide core and cladding, respectively. The spatial components ($x$, $y$, $z$) of the electric field $\vec{E}(x, y, z)$ at both core–cladding and cladding–air interfaces are obtained using FEM simulations. As the typical amplitude of sidewall roughness is much less than the waveguide width, it is justified to use the modal field profile of the ideal, smooth waveguide, derived from FEM, as a first-order approximation for $\vec{E}(x, y, z)$ at the interface, thus simplifying the scattering analysis. This approach, however, discounts the influence of high refractive index contrast and internal radiation within the core. To account for these limitations, we employ dyadic Green's functions for a single-layer medium, as described in \cite{SIBarwicz:05}.

Interface roughness is mathematically represented by a spatial profile function $f(z)$ along the propagation direction. Its statistical properties are described by the autocorrelation function, $R(u) = \langle f(z)f(z+u)\rangle$, commonly modeled as exponentially decaying, $R(u) \approx \sigma^2 \exp(-|u|/L_c)$, where $\sigma^2$ is the variance and $L_c$ the correlation length of the roughness \cite{SIBarwicz:05,SIBauters:11,SIBauters:13}. Note that $\sigma^2$ and $L_c$ typically differ for the vertical sidewalls and horizontal surfaces; additionally, it is assumed that the roughness statistics between distinct boundaries are mutually uncorrelated, allowing each interface to be treated independently in the scattering analysis. The spatial frequency content of the roughness is quantified by the Fourier transform of $R(u)$: $\widetilde{R}(\xi) = \mathscr{F}[R(u)] \approx \frac{2\sigma^2 L_c}{1 + L_c^2 \xi^2}$, where this expression characterizes how various spatial frequencies $\xi$ contribute to the overall surface profile and to scattering \cite{SIBauters:11}.

The ensemble-averaged far-field radiation per unit length ($P_{\mathrm{rad}}/L$) that emerges because of surface roughness can be computed via an array factor formalism:

\begin{equation}
    \frac{P_{\mathrm{rad}}}{L} = \int_{0}^{2\pi} \int_{0}^{\pi} (\vec{S} \cdot \hat{r})\, \widetilde{R}(\beta - n_{\mathrm{clad}} \hat{r} \cdot \hat{z})\, r^2 \sin\theta\, d\theta\, d\phi
\end{equation}
where $(\vec{S} \cdot \hat{r})$ is the component of the Poynting vector in the radiation direction, encapsulating the combined effect of modal overlap and dyadic Green's function response \cite{SIBarwicz:05} and and $\beta$ is the propagation constant of the propagating mode. The scattering loss per unit length obtained from FEM simulations for our low-confinement waveguides with core thicknesses of $50~\mathrm{nm}$ and $100~\mathrm{nm}$ at an optical wavelength of $780~\mathrm{nm}$ is shown in Supplementary Figure~\ref{Fig_Loss-reponsivity}. The calculations assume a roughness variance of $\sigma^2 = 5~\mathrm{nm}^2$ and a correlation length of $L_c = 50~\mathrm{nm}$.

\subsection{2D loss model}

The two-dimensional scattering loss formalism used here is a normalized variant \cite{CorZan:24} of the Payne–Lacey model \cite{SILacey:90}. The original Payne–Lacey model is a perturbative, semi-analytical 2D treatment of roughness-induced scattering in slab and strip waveguides. The normalized formulation removes geometry-dependent modal overlap factors, yielding a universal description of roughness-induced loss applicable across different waveguide cross sections. The normalized sidewall (or top/bottom) scattering loss coefficient is given by
\begin{equation}
\alpha^{\mathrm{norm}}_{\mathrm{sides/top/bottom}}(\lambda)
=
\left(
n_{\mathrm{core}}^{2}-n_{\mathrm{clad}
)^{2}}
\right)^{2}
\frac{k_{0}^{3}}{4\pi n_{\mathrm{core}}}
S_{\mathrm{sides/top/bottom}}.
\label{norm_alpha}
\end{equation}

Here, $n_{\mathrm{core}}$ and $n_{\mathrm{clad}}$ denote the refractive indices of the silicon nitride core and silicon dioxide cladding, respectively, $k_{0}=\frac{2\pi}{\lambda}$ and $S_{\mathrm{sides/top/bottom}}$ is the corresponding geometric correction factor, defined as

\begin{align}
S_{\mathrm{sides/top/bottom}}
&=
\frac{\sqrt{2}\sigma^{2} L_{c} \,\pi}{D}
\left(
\sqrt{D + 1 - L_{c}^{2}\!\left(\beta^{2}-n_{\mathrm{clad}}^{2}k_{0}^{2}\right)}
\right)
\\[4pt]
D
&=
\sqrt{4\beta^{2}L_{c}^{2}+\left[1-L_{c}^{2}\!\left(\beta^{2}-n_{\mathrm{clad}}^{2}k_{0}^{2}\right)
\right]^{2}}.
\notag
\end{align}

In this above expression, $\sigma$ and $L_c$ denote the RMS amplitude and correlation length of the sidewall and top/bottom roughness, respectively, and $\beta$ is the propagation constant of the propagating mode. The total scattering loss for mode $n$ is obtained by multiplying the normalized sidewall coefficient by the modal overlap ($\phi_n$) with the sidewall (or top/bottom):
\begin{equation}
\alpha^{(n)}_{\mathrm{total}} = \alpha^{\mathrm{norm}}_{\mathrm{sides}} \phi_n.
\end{equation}

For the initial scattering-loss assessment, we employ a three-dimensional loss model to evaluate waveguides with high aspect ratios and core thicknesses of $50~\mathrm{nm}$. The scattering loss per unit length ($\alpha_{\mathrm{scat}}/L$) is computed using electric-field distributions obtained from FEM simulations, as shown in Supplementary Figure~\ref{Fig_Loss-reponsivity}f. 

The results indicate that $\alpha_{\mathrm{scat}}/L$ increases with core thickness and decreases with core width. While a higher aspect ratio reduces scattering loss, it simultaneously necessitates a thicker cladding and greater core-to-actuator separation to accommodate the expanded optical mode profile. This increased separation degrades the electro-optic responsivity, requiring a longer phase-shifter length and thereby increasing the total propagation loss. Consequently, the waveguide geometry must be optimized using a combined figure of merit such as $V_{\pi}\alpha$ (equivalently, the voltage–length–loss product, VLP).

To reduce computational cost during the optimization procedure, we adopt the 2D scattering loss model and make two simplifying approximations. First, we treat the sidewall scattering loss as the dominant scattering contribution and neglect top/bottom surface scattering contribution, due to substantially lower RMS roughness of top/bottom surface. Second, we approximate the modal overlap factor with the sidewalls as constant, $\phi_n \approx 1$, throughout the design sweep, that is $\alpha^{(n)}_{\mathrm{total}} = \alpha^{\mathrm{norm}}_{\mathrm{sides}}$. This approximation is well justified in our geometry because the waveguide width is held fixed at $50~\mathrm{nm}$, resulting in negligible variation in the transverse field profile and hence in the sidewall overlap across the parameter space explored. Moreover, since the scattering loss is known to decrease monotonically with increasing waveguide width, fixing $\phi_n \approx 1$ does not shift the location of the global optimum in the remaining design-variable space. Under this framework, the total propagation loss is expressed as $\alpha = \alpha^{(n)}_{\mathrm{total}} + \alpha_{\mathrm{abs}}$, where $\alpha_{\mathrm{abs}}$ is the material absorption loss, and the figure of merit becomes $\mathrm{VLP} = \alpha/L \cdot V_{\pi}L$. Since $\alpha^{\mathrm{norm}}_{\mathrm{sides}}$ is calculated for a single sidewall, the VLP value is doubled to account for both sidewalls. Throughout the optimization, we use the experimentally determined piezo-optic coefficients for silicon nitride: $p_{11} = -0.125$ and $p_{12} = 0.047$.

\begin{figure}[ht!]
\centering
\includegraphics[width=\linewidth]{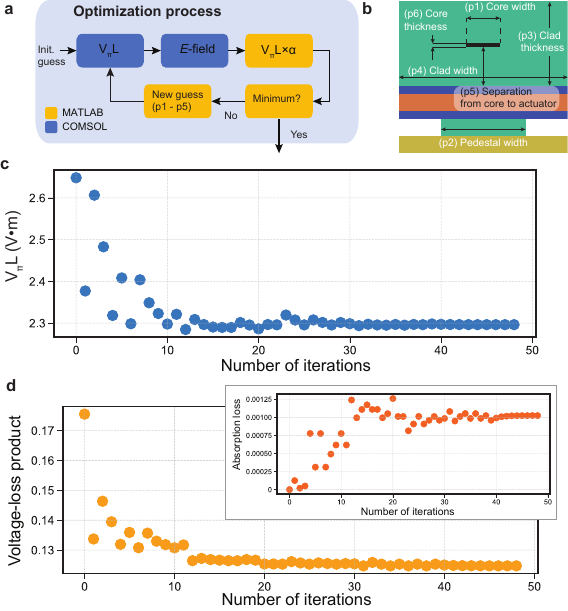}
\caption{\textbf{Optimization framework and convergence metrics for waveguide geometry.}\\
\textbf{(a)} Flowchart outlining the implementation of the optimization algorithm, highlighting computations performed in MATLAB (yellow blocks) and analysis stages in COMSOL Multiphysics (dark blue blocks). \textbf{(b)} Set of parameters used in the optimization of the waveguide geometry. \textbf{c, d} Evolution of the $V_{\pi} L$ and VLP metrics over the course of the optimization, both converging to a minimum after approximately 50 iterations. The inset in d shows the absorption loss contributed by the actuator’s top metal layer at different iterations. The core thickness is held fixed at $50~\mathrm{nm}$ throughout the optimization.}
\label{Optimization}
\end{figure}

\section{Optimal waveguide geometry for minimum VLP with silicon nitride piezo-optic coefficients $\mathbf{p_{11}=-0.125}$ and $\mathbf{p_{12}=0.047}$}
\label{VLP_optimize}

We utilized the gradient descent (patternsearch) optimization algorithm from the MATLAB Global Optimization Toolbox to minimize VLP by optimizing the geometry. The algorithm can optimize the high-dimensional parameter space while simultaneously managing various upper and lower bound constraints. In our specific application, these constraints are the maximum and minimum dimensions of the core and cladding, the positioning of the core within the waveguide, and the width of the pedestal, as illustrated in Supplementary Figure~\ref{Optimization}b. The core width ranges from $1~\mathrm{\mu m}$ to $4~\mathrm{\mu m}$, the core thickness is fixed at $50~\mathrm{nm}$, the cladding width ranges from $7~\mathrm{\mu m}$ to $18~\mathrm{\mu m}$, the cladding thickness ranges from $7~\mathrm{\mu m}$ to $14~\mathrm{\mu m}$, the pedestal thickness from $3~\mathrm{\mu m}$ to $4~\mathrm{\mu m}$, and the height of the core ($h$) from the top Al electrode of the actuator ranges from $2~\mathrm{\mu m}$ to $6~\mathrm{\mu m}$.

The simulation process is designed as a feedback loop, as illustrated in Supplementary Figure~\ref{Optimization}a. In each loop, the pattern search algorithm in MATLAB modifies the parameter values, and COMSOL Multiphysics (connected to MATLAB via LiveLink) solves for the effective refractive index, loss values, and the VLP value. The robustness of the optimization algorithm is evaluated by subjecting it to various initial parameter values. The core width is initialized at several values, including $1~\mathrm{\mu m}$, $1.6~\mathrm{\mu m}$, and $2~\mathrm{\mu m}$. Regardless of the initial core width, the algorithm consistently converges to the same set of waveguide dimensions. For a given set of parameter ranges, the VLP value for a core thickness of $50~\mathrm{nm}$ converges after 48 iterations (as shown in Supplementary Figure~\ref{Optimization}d), with a final $V_{\pi}L$ of 2.3 $\mathrm{V}\cdot \mathrm{m}$. The resulting core width is $1~\mathrm{\mu m}$, with cladding thickness and width of $11.6~\mathrm{\mu m}$ and $12.2~\mathrm{\mu m}$, respectively. It is evident from the results that optimiation algorith tries to reduce the silicon nitride contribution to $V_{\pi} L$ by reducing the width of the waveguide to lower bound. All scattering loss calculations assume a surface roughness variance ($\sigma^2$) of $5~\mathrm{nm}^2$ for the sidewall and $0.001~\mathrm{nm}^2$ for the top/bottom surfaces, and a correlation length ($L_c$) of $50~\mathrm{nm}$.

\section{Sources of excess propagation loss in actuator integrated waveguides}

In unetched, low-confinement silicon nitride waveguides without an actuator, the dominant loss mechanism is scattering at the core--cladding interface, yielding a propagation loss of 2.5--3~dB/m for a 1.8~m spiral. In contrast, etched spiral waveguides integrated with a bottom piezoelectric actuator exhibit a propagation loss of 7.2--8.3~dB/m over an 8.2~cm spiral, indicating an excess loss of approximately 4--6~dB/m. This excess loss arises from three distinct mechanisms, each analyzed separately below.

\begin{figure}[ht!]
\centering
\includegraphics[width=\linewidth]{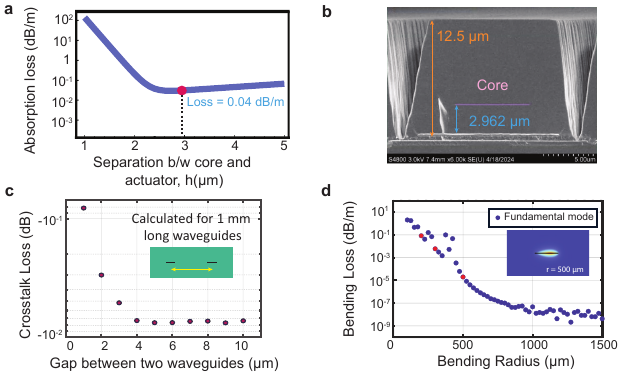}
\caption{\textbf{FEM simulations of absorption loss, inter-waveguide crosstalk, 
and bending loss in the low-confinement silicon nitride platform.}\\
\textbf{(a)} Simulated absorption loss as a function of metal--core separation. 
The fabricated devices have a separation of ${\sim}2.96~\mathrm{\mu m}$, 
yielding an absorption loss of less than $0.04~\mathrm{dB/m}$.
\textbf{(b)} SEM image of the cross-section of the waveguide with separation between core and actuator labeled as $2.962~\mathrm{\mu m}$.
\textbf{(c)} Simulated inter-waveguide crosstalk as a function of waveguide 
spacing for $1~\mathrm{mm}$-long waveguides. Crosstalk becomes negligible 
beyond ${\sim}10~\mathrm{\mu m}$; the current waveguide pitch of 
${\sim}22.5~\mathrm{\mu m}$ provides a conservative margin well above 
this threshold.
\textbf{(d)} Simulated bending loss for the fundamental mode as a function 
of bend radius. Red points indicate the bending loss at the three fabricated 
minimum bend radii of $150~\mathrm{\mu m}$, $300~\mathrm{\mu m}$, and 
$500~\mathrm{\mu m}$, confirming that bending loss is negligible at 
$300~\mathrm{\mu m}$ and above, and increases rapidly below this value.}
\label{Absorption_loss}
\end{figure}

\subsection{Absorption loss from the metal electrodes}

The metal electrodes of the piezoelectric actuator stack introduce optical 
absorption loss through evanescent overlap of the guided mode with the metal 
layers. To balance metal-induced absorption loss against piezoelectric coupling 
efficiency, we use FEM simulations to compute both the voltage--length product 
$V_\pi \cdot L$ and the absorption-loss-limited voltage--loss product 
$V_\pi \cdot \alpha_\mathrm{abs}$ as functions of the waveguide core height 
$h$ relative to the actuator base, shown in Supplementary 
Figure~\ref{Fig_Loss-reponsivity}e. The two curves exhibit opposing trends: decreasing 
$h$ improves piezoelectric coupling (lower $V_\pi \cdot L$) but increases metal 
absorption (higher $\alpha_\mathrm{abs}$), defining a clear 
trade-off. The buried oxide height that minimizes the overall 
$\mathrm{VLP} = V_\pi \cdot L \times \alpha_\mathrm{abs}$, explicitly accounting for both 
contributions, corresponds to a metal--core separation of 
${\sim}2.5~\mathrm{\mu m}$. Our fabricated devices have a metal--core 
separation of approximately $2.96~\mathrm{\mu m}$, close to, though slightly 
above, this simulated optimum.
Supplementary Figure~\ref{Absorption_loss}a shows the absorption loss as a 
function of metal--core separation, confirming that our fabricated separation 
of ${\sim}2.96~\mathrm{\mu m}$ yields less than $0.05~\mathrm{dB/m}$ of 
absorption loss. Supplementary Figure~\ref{Absorption_loss}c shows the 
inter-waveguide crosstalk as a function of waveguide spacing for 
$1~\mathrm{mm}$-long waveguides, confirming that crosstalk becomes negligible 
beyond ${\sim}10~\mathrm{\mu m}$; our current pitch of ${\sim}22.5~\mathrm{\mu m}$ 
provides a conservative margin well above this threshold. Supplementary 
Figure~\ref{Absorption_loss}d shows the simulated bending loss for the 
fundamental mode as a function of bend radius; the highlighted points at 
$150~\mathrm{\mu m}$, $300~\mathrm{\mu m}$, and $500~\mathrm{\mu m}$ bend 
radii confirm that bending loss is negligible for our fabricated devices at 
$300~\mathrm{\mu m}$ and above, and increases rapidly below this value.

\subsection{Scattering loss from the cladding–metal interface}

\begin{figure}[ht!]
\centering
\includegraphics[width=\linewidth]{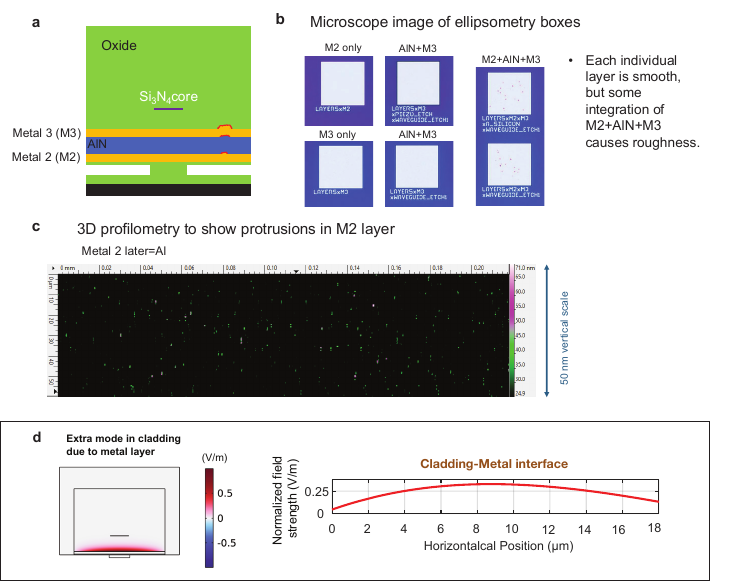}

\caption{\textbf{Characterization of metal-layer surface roughness at the 
cladding--metal interface.}\\
\textbf{(a)} Schematic cross-section illustrating nanoscale protrusions at 
the M2--AlN interface that extend into the oxide cladding, introducing 
surface roughness and acting as localized sources of optical scattering loss.
\textbf{(b)} Optical microscope images of ellipsometry test structures showing 
individual deposition layers (M2, AlN, M3) and their combined M2+AlN+M3 stack. 
While each individual layer appears smooth, certain regions of the combined 
stack exhibit small protrusions that introduce additional surface roughness.
\textbf{(c)} Three-dimensional profilometry of the M2 layer over a 
$0.22~\mathrm{mm} \times 60~\mathrm{\mu m}$ scan area, showing an RMS 
roughness of $8.4~\mathrm{nm}$. Most surface peaks fall in the 
$30$--$60~\mathrm{nm}$ range, and roughness values may be even higher in 
the M3 layer.  \textbf{(d)} FEM analysis revealing a cladding--metal interface mode with high field strength at the metal boundary, representing an additional potential source of scattering loss due to metal protrusions extending into the cladding.}
\label{M2_AlN_M3}
\end{figure}

In our low-confinement geometry, the guided mode resides primarily in the 
silicon nitride core and the surrounding oxide cladding, with direct overlap 
with the metal electrodes minimized by vertically separating the actuator 
stack from the waveguide core. However, imperfections in metal deposition 
can introduce nanoscale protrusions at the metal--oxide interface, which 
locally act as sources of optical scattering loss.

\begin{figure}[ht!]
\centering
\includegraphics[width=\linewidth]{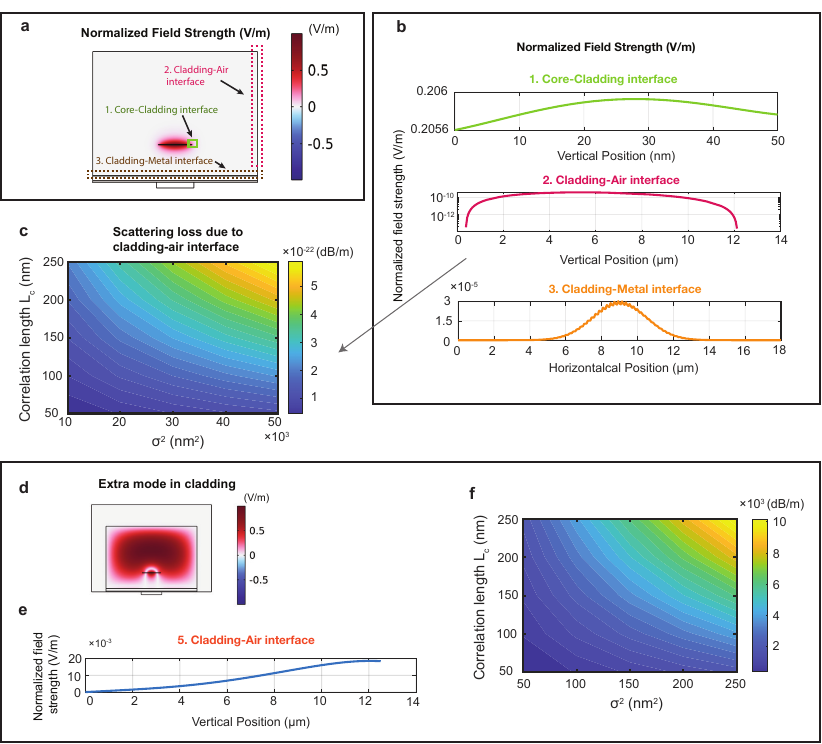}
\caption{\textbf{Scattering loss analysis at the cladding--air interface.}\\
\textbf{(a)} Cross-sectional mode profile of the low-confinement silicon 
nitride waveguide, showing the guided field distribution across the waveguide 
stack and the locations of the core--cladding, cladding--metal, and 
cladding--air interfaces.
\textbf{(b)} Normalized field strength at the core--cladding 
(${\sim}10^{-1}~\mathrm{V/m}$), cladding--air
(${\sim}10^{-10}~\mathrm{V/m}$), and cladding--metal
(${\sim}10^{-5}~\mathrm{V/m}$) interfaces, illustrating the strong 
attenuation of the evanescent field by the $17$--$20~\mathrm{\mu m}$ thick 
oxide cladding.
\textbf{(c)} Calculated scattering loss from the cladding--air interface as a 
function of roughness parameters (correlation length and RMS roughness), 
computed using the model of Ref.~\cite{SIBauters:11}. The scattering loss 
remains of the order of $10^{-22}~\mathrm{dB/m}$ across all roughness 
parameters considered, confirming that this interface contributes negligibly 
to the total propagation loss. \textbf{(d,e)} The etched waveguide geometry supports a cladding-guided 
mode within the $9.5~\mu\mathrm{m}$ thick oxide cladding above the core, absent 
in the unetched waveguide, with its corresponding cladding-air sidewall field profile. \textbf{(f)} Analytical calculations of the scattering loss for this mode yield values on the order of $10^{3}~\mathrm{dB/cm}$, indicating that any light coupled into this mode is effectively radiated away.}
\label{Clad_Air}
\end{figure}

Supplementary Figure~\ref{M2_AlN_M3}a shows a schematic illustrating 
protrusions at the M2--AlN interface that extend into the oxide cladding, 
introducing surface roughness and acting as localized sources of scattering 
loss. In the microscope image of the ellipsometry test structures 
(Supplementary Figure~\ref{M2_AlN_M3}b), each individual layer appears 
smooth; however, certain regions of the combined Metal 2 (M2) + AlN layer + Metal 3(M3) stack exhibit 
small protrusions that introduce additional roughness. These protrusions are 
directly visible in the 3D profilometry of the M2 layer 
(Supplementary Figure~\ref{M2_AlN_M3}c), obtained over a 
$0.22~\mathrm{mm} \times 60~\mathrm{\mu m}$ scan area, with an RMS 
roughness of $8.4~\mathrm{nm}$. Most peaks in the profilometry fall in the 
$30$--$60~\mathrm{nm}$ range, and these values may be even higher in the 
M3 layer. Given that the modal field amplitude at the cladding--metal 
interface is of the order of $10^{-5}~\mathrm{V/m}$ 
(Supplementary Figure~\ref{Clad_Air}b), this interface represents 
a significant source of scattering loss. Additionally, FEM analysis reveals a cladding–metal interface mode with relatively high field strength at the metal boundary. Interaction of this mode with metal protrusions extending into the cladding represents an additional potential source of scattering loss, as shown in Fig. \ref{M2_AlN_M3}d.

\subsection{Scattering loss from the cladding--air interface}

The etching process that defines the released sections exposes the outer oxide 
cladding surface to air, creating a cladding--air interface that could in 
principle contribute to scattering loss. To assess this contribution, we 
compute the normalized field strength at the core--cladding, cladding--air, 
and cladding--metal interfaces, shown in Supplementary 
Figure~\ref{Clad_Air}b, which are of the order of $10^{-1}~\mathrm{V/m}$, 
$10^{-10}~\mathrm{V/m}$, and $10^{-5}~\mathrm{V/m}$, respectively. The 
cross-sectional mode profile in Supplementary Figure~\ref{Clad_Air}a 
illustrates how the guided field decays across the $17$--$20~\mathrm{\mu m}$ 
thick oxide cladding before reaching the cladding--air interface, resulting in 
a field amplitude seven orders of magnitude smaller than at the 
core--cladding interface. Using the scattering loss model of 
Ref.~\cite{SIBauters:11}, the scattering loss contribution from the 
cladding--air interface is shown in Supplementary Figure~\ref{Clad_Air}c as a 
function of roughness parameters. The calculated scattering loss is of the 
order of $10^{-22}~\mathrm{dB/m}$, which is negligible across all roughness 
parameters considered. This negligible contribution arises because the field 
amplitude at the cladding--air interface is strongly attenuated by the thick 
oxide cladding, resulting in minimal evanescent interaction with the outer 
surface. Additionally, the etched waveguide geometry supports a cladding-guided mode within the 9.5 µm thick oxide cladding above the core, which is absent in the unetched waveguide, as shown in Fig \ref{Clad_Air}d, with the corresponding cladding-air sidewall field profile presented in Fig \ref{Clad_Air}e. Analytical calculations of the scattering loss associated with this mode yield values on the order of $10^{-3}$ dB/cm (Fig \ref{Clad_Air}f), indicating that any light coupled into this mode is effectively radiated away.

We therefore conclude that the dominant source of excess loss is scattering at the cladding–metal interface, where a modal field interacts with surface roughness introduced by metal protrusions in the actuator stack, with potential additional contributions from higher-order cladding modes supported by the etched geometry and cladding-metal interface mode.

\end{document}